\documentclass[reprint,showpacs,preprintnumbers,amssymb,nofootinbib,aps,prd]{revtex4-1}

\usepackage{graphicx,rotate}
\usepackage{amsmath,amsfonts,amssymb}
\usepackage{epsfig,epsf}
\usepackage{array}
\usepackage[usenames,dvipsnames]{color}
\usepackage{hyperref,bm}
\usepackage{dcolumn,subfig}

\usepackage[normalem]{ulem}

\definecolor{darkgreen}{cmyk}{1,0,1,0.4}

\definecolor{pink}{cmyk}{0.4,1,0.3,0}

\def\com2#1{\textcolor{red}{\it{#1}}}

\def\bar {\overline}

\def\be {\begin{equation}}
\def\ee {\end{equation}}
\def\beq {\begin{equation}}
\def\eeq {\end{equation}}
\def\bea {\begin{eqnarray}}
\def\eea {\end{eqnarray}}

\def\bra {\langle}
\def\ket {\rangle}

\def\beq{\begin{equation}}
\def\eeq{\end{equation}}
\def\barr{\begin{array}}
\def\earr{\end{array}}

\newcommand{\nn}{\nonumber}
 
\begin{document}

\title{Optimal observable analysis for the decay $b \to s$ plus missing energy}

\author{Zaineb Calcuttawala}
\email{zaineb.calcuttawala@gmail.com}
\affiliation{Department of Physics, University of Calcutta, 92 Acharya Prafulla Chandra Road, Kolkata 700009, India}

\author{Anirban Kundu}
\email{anirban.kundu.cu@gmail.com}
\affiliation{Department of Physics, University of Calcutta, 92 Acharya Prafulla Chandra Road, Kolkata 700009, India}

\author{Soumitra Nandi}
\email{soumitra.nandi@gmail.com}
\affiliation{Department of Physics, Indian Institute of Technology, Guwahati 781039, India}

\author{Sunando Kumar Patra}
\email{sunando.patra@gmail.com}
\affiliation{Department of Physics, Indian Institute of Technology, Guwahati 781039, India}

\begin{abstract}

The decay $b\to s\nu\bar\nu$ has been a neglected sibling of $b\to s\ell^+\ell^-$ because neutrinos pass 
undetected and hence the process offers lesser number of observables. We show how the decay 
$b\to s~+$~invisible(s) can shed light, even with a limited number of observables,
on possible new physics beyond the Standard Model and also show,
quantitatively, the reach of future $B$ factories like SuperBelle to uncover such new physics. Depending 
on the operator structure of new physics, different channels may act as the best possible probe. We show,
using the Optimal Observable technique, how almost the entire parameter space allowed till now can 
successfully be probed at a high luminosity $B$ factory. 

\end{abstract}

\maketitle


\section{Introduction}

While the semileptonic decay $B\to K^{(*)}\ell^+\ell^-$ mediated by the flavor-changing neutral 
current (FCNC) transition $b\to s$ has received a lot of attention as a sensitive probe of new physics 
(NP) beyond the Standard Model (SM), much less light has been shed on the analogous process 
$b\to s\nu\bar{\nu}$; either the exclusive channels $B\to K^{(*)}\nu\bar\nu$ or the semi-inclusive one 
$B\to X_s\nu\bar\nu$. There are three main reasons for this. First, the decay is yet to be observed; 
there is only an upper limit to the branching ratio (BR) of such processes \cite{pdg-2016,
Grygier:2017tzo}. This is not 
unexpected considering that the experimental sensitivity is about one order of magnitude above 
the SM predictions. Second, 
the number of observables are less than the processes involving charged leptons, because the 
neutrinos escape undetected. Third, the theoretical uncertainties like those coming from the 
form factors are more serious than relatively cleaner channels like $K\to \pi\nu\bar\nu$. 

One can pose counter-arguments too. For example, the Belle upgrade with a much enhanced 
integrated luminosity (or any other future 
$e^+e^-$ $B$ factory) will almost definitely observe this process even if there is no NP involved. 
The limited number of observables can actually make the analysis cleaner. 
As we will show quantitatively, one can successfully use these few observables not only to differentiate 
some well-motivated NP models from the SM but also to have a glimpse of the possible 
operator structure 
of those models. This remains true even when one takes into account all the theoretical uncertainties like the 
form factors, elements of the Cabibbo-Kobayashi-Maskawa (CKM) matrix, the running quark masses, 
or the higher-order corrections. 

The conception that whatever may be inferred from the neutrino channels can also be inferred 
in a much cleaner way involving charged lepton final states, because of the $SU(2)_L$ conjugate 
nature of the corresponding operators, is also not entirely correct. Consider, for example, an
$SU(2)_L$ singlet  current 
of the form $\epsilon_{ab} \bar{L}_L^a \gamma^\mu Q_L^b$, involving both quark and lepton doublets.
that couples to a vector leptoquark. The charged lepton final states, obviously, come only from 
the anomalous top decays and not from $B$ decays. Thus, the neutrino channels are worthy to be studied 
on their own right. There may be other light invisible particles in the final state (we will show an example later) 
which have nothing to do with charged leptons. As neutrinos or other such invisibles go undetected, 
this channel offers an effective probe for such models. Thus, one must treat the decay channel 
$b\to s~+$~invisible(s) as an independent source of information from $b\to s\ell^+\ell^-$, although 
there can be correlations in some beyond-SM models.  

As an example of what we have just said, let us note that apart from neutrinos 
coming from non-SM operators, the 
FCNC transition $b\to s$ can involve light invisible scalars in the final state. If they are singlet under 
all the SM gauge groups, they can very well be a candidate for cold dark matter (DM). 
Although there are strong constraints on such light DM particles from the direct detection experiments like 
LUX, XENON or PANDAX
\cite{dm-limits}, one may avoid them if the DM is nonthermal in origin. The only limit comes from 
the invisible decay width of the Higgs boson at the Large 
Hadron Collider (LHC), which one can easily keep within the tolerable limit of $\sim 10\%$ if the 
Higgs-DM coupling is small.
Thus, an analysis of the decay $b\to s~+$ invisibles will also act as a complementary 
probe to the DM direct search.

We will use the Optimal Observable (OO) technique for the analysis. The OO technique helps one to 
identify observables where the NP can be differentiated from the SM with highest confidence level. 
Of course, which observables are to be used depends on which part of the parameter space the NP 
falls. While this technique has been more widely used for collider studies \cite{gunion,
Atwood:1991ka}, Ref.\ \cite{S3} 
shows how this can be applied to semileptonic $B$ decays as well. Note that in the absence of any data, 
one must work only with statistical uncertainties. The systematic uncertainties will somewhat relax the 
reaches and the confidence levels. 

The OO technique not only shows the regions 
of the parameter space for NP where differentiation from the SM will be easy but also the variables that 
one should look at to have such a successful differentiation. In other words, a simultaneous study of 
all the relevant observables can effectively pin down the region of the parameter space where any
beyond-SM physics may lie. In the next Section, we will provide a sketchy 
discussion of the OO technique. In Section
III, we discuss two most popular NP models; one with neutrinos in the final state but the effective operator
basis augmented by some NP operators; and the second with light scalars in the final state along with 
SM neutrinos. Sections IV and V discuss our results for these two models respectively. 
We summarize and conclude in Section VI.

\section{The optimal observable technique}

This section is rather sketchy and follows the notation of Ref.\ \cite{gunion,Atwood:1991ka}
Suppose there is an observable $O$ which depends on the variable $\phi$ as
\be
O(\phi) = \sum_i c_i f_i(\phi)\,,
\ee
where $c_i$s are model-dependent coefficients, like the Wilson coefficients (WC), and $f_i(\phi)$ are known 
functions of $\phi$. For our case, $\phi$ can be identified with the momentum transfer (to the 
invisible particles) squared, $q^2=(p_B - p_{K^{(*)}})^2$, where $p_a$ denotes the four-momentum of the 
particle $a$.  To get $c_i$, one can fold with weighting functions $w_i(\phi)$ such that 
\be
\int w_i(\phi) O(\phi) \, d\phi = c_i\,.
\ee
There happens to be a unique choice of $w_i(\phi)$ such that the statistical error 
in $c_i$s are minimized. For this choice, the covariance matrix $V$, defined as 
\be
V_{ij} \propto \int w_i(\phi) w_j(\phi) O(\phi)\, d\phi 
\ee
is at a stationary point with respect to the variation of $\phi$:
$\delta V_{ij} = 0$. 
This happens if we choose
\be
w_i(\phi) = \frac{ \sum_j X_{ij} f_j(\phi) } { O(\phi)}\,,
\ee
where
\be
X_{ij} = (M^{-1})_{ij}\,,\ \ 
M_{ij} = \int \frac{f_i(\phi)f_j(\phi) }{O(\phi)} \, d\phi\,. \label{mijdef}
\ee
In this case,
\be
c_i = \sum_j X_{ij} I_j = \sum_j (M^{-1})_{ij} I_j\,, \ \ 
I_j = \int f_j(\phi)\, d\phi\,.
\ee

For only this choice of weighting functions, the covariance matrix is 
\be
V_{ij} = \bra \Delta c_i \Delta c_j\ket = \frac{ (M^{-1})_{ij} \sigma_T}{N}\,,
\label{vijdef}
\ee
where $\sigma_T = \int O(\phi)\, d\phi$. (If $O(q^2) = d\Gamma/dq^2$, $\sigma_T=\Gamma$.) 
$N$ is the total number of events, given by the integrated cross-section times total luminosity times 
the efficiencies. This result holds even if there are applied cuts. 
The minimum of statistical 
uncertainty in the extraction of a parameter gives the maximum significance of that 
parameter over the others. Therefore,
using this technique we can test the significance of a specific NP model over the other models,
including the SM. In other words, given the data, one can say with what significance some observable 
may differentiate a particular type of NP from the SM. This significance, as one should emphasize here,
depends on the observable chosen, on the parameters of the NP model, and on the integrated 
luminosity, all of which are intuitively obvious.

As an example, suppose one is looking at the branching fractions of a $B$ meson to 
several final states. For the final state $f$, the branching fraction can be expressed as 
\be
{\cal B}(B\to f)^{\rm exp} = \frac{1}{\Gamma} \, \int \frac{d\Gamma}{dq^2}\, dq^2\,,
\ee
where $\Gamma$ is the total decay width.
The statistical uncertainties in $c_i$s extracted from the branching fractions can be written as \cite{S3} 
\be
|\delta c_i| = \sqrt{\frac{X_{ii} {{\cal B}(B\to f)}^{\rm exp}}{N}} = \sqrt{\frac{X_{ii}}{\sigma_P {\cal L}_{\rm eff}}}\,.
\label{eq5}
\ee
As given in Eq.\ (\ref{eq5}),
the errors are also related to the total production cross section $\sigma_P$ ( = $\sigma_{B \to f}/{\cal B}(B \to f)$),
and the effective luminosity ${\cal L}_{\rm eff} = {\cal L}_{\rm int} \epsilon_s$, where
${\cal L}_{\rm int}$ and $\epsilon_s$ are the integrated luminosity and reconstruction efficiency respectively.

%
%

When the number of nonzero NP parameters is small, the analysis can also be done by 
defining a quantity analogous to $\chi^2$, such as 
\begin{align}\label{chidef}
 \chi^2 &= \sum_{i,j} (c_i - c_i^0) (c_j - c_j^0) V_{ij}^{-1}.
\end{align}
The $c_i^0$s are called the seed values, which can be considered as model inputs.
Thus, they are the values of $c_i$s with the parameter values chosen for the reference model.
$V_{ij}$s are defined in Eq.\ (\ref{vijdef}). 

As mentioned earlier, our goal is to study the decay $b\to s~+$ invisible(s), which includes the 
exclusive modes like $B \to K^{(*)} \nu{\bar\nu}$ 
\footnote{Technically, this is not fully exclusive, as we do not care about the flavor of the neutrinos.}. 
In such decays, the major sources of uncertainties 
are the hadronic form-factors, like $F^{B\to K^{(*)}}(q^2)$. 
Thus, it is important to differentiate between SM and any possible NP taking into account all these
uncertainties, and check whether 
the future experimental statistics will allow a clear separation of the two. 
Let us now explain briefly how to use a $\chi^2$ statistic test to pinpoint such a differentiation. 

There are a few things one should take note of while interpreting the results of the OO technique. 

(1) A regular $\chi^2$ statistic is a function of parameters of a model and `measures' the deviation of those 
from observed values of some experimental quantities. The one we need to concoct, however, should take 
one specific model ({\em e.g.} SM) as reference, in place of experimental results, and should be a function 
of some parameters, each set of values of which indicates one comparison-worthy model (e.g. NP). 
For example, if we have a set of new operators $O_i$ with $c_i$ being the corresponding WCs, 
$c_i=0\, \forall\, i$ is the SM and any other set is a NP model that may be compared with the SM. 
The $\chi^2$ should also be a measure of `separation' (deviation) between any NP model and 
the reference one. In the rest of this analysis, the reference-model will always be SM. By construction, 
we ensure that $\chi^2|_{SM} = 0$ and $\chi^2 = n^2$ denotes a separation of $n~\sigma$ from the SM.

(2) Projections of the constant $\chi^2 = 1$ contours on each parameter axis will give us the 
corresponding $\delta c_i$s. Following the point above, the constructed $\chi^2$ has no measurements 
or data points in it and thus the $\delta c_i$s obtained are not statistical uncertainties on the $c_i$s; 
it is not even the predictions for them. These are uncertainties of the reference model, or, in other words,
a measure of the region in parameter space where the reference model is indistinguishable from models parametrized by surrounding points in the parameter space. 
Thus, points on the $1\sigma$ contour 
parametrize models that can be distinguished from the SM at $1\sigma$ level only.

(3) When varying the $\chi^2$ over the allowed parameter space, $V_{i j}^{-1}$ also depends on the 
parameter values through $O(\phi)$, which comes in the denominator of $M_{i j}$ (Eq.\ (\ref{mijdef})). 
This is known as the seed dependence.

(4) The covariance matrix $V_{i j}$ as well as the $\chi^2$ are obtained using the central values of all 
the parameters. So any separation obtained after the analysis, though qualitatively correct, has to be 
modified after inclusion of the SM errors. 

(5) If considered, theoretical uncertainties in $O(\phi)$ will in turn introduce an uncertainty in the $\chi^2$. 
In other words, in presence of the uncertainties, the $n~\sigma$ contours will become bands of nonzero 
width in the parametric space.

For completeness, we will also provide the decay rate distributions of the processes under consideration, 
and test their usefulness in differentiating the NP models from the SM.

\section{New physics models and observables}

\subsection{Only neutrinos as invisible}

The first NP model treats neutrinos as the only carrier of missing energy in $b\to s$ decays. We will also 
take, for simplicity, not only no lepton flavor violation (LFV) but also lepton flavor universality (LFU). This 
means all the three flavors of $\nu\bar\nu$ pairs are produced in equal number even by the NP operators,
and there are no $\nu_i\bar\nu_j$ final states with $i\not=j$. Note that both these assumptions can be 
violated in specific NP models.

The effective Hamiltonian for $b\to s \nu_i\bar\nu_i$ can be written as 
\be
{\cal H}_{\rm eff} = \frac{4G_F}{\sqrt{2}} V_{tb} V^*_{ts} \left[ C_{SM} O_{SM} + C_{V_1}O_{V_1} + C_{V_2} O_{V_2}\right]\,,
\label{h-eff}
\ee
where
\bea
O_{SM} = O_{V_1} &=& \left( \bar{s}_L \gamma^\mu b_L\right)\left(\bar\nu_{iL}\gamma_\mu \nu_{iL}\right)\,,
\nonumber\\
O_{V_2} &=& \left( \bar{s}_R \gamma^\mu b_R\right)\left(\bar\nu_{iL}\gamma_\mu \nu_{iL}\right)\,.
\label{ops}
\eea
Note that $O_{SM}$ and $O_{V_1}$ are identical only with our assumption of LFU and no LFV.  
NP with LFU can mean, in an extreme case, that only one flavor of neutrino will be present;
with LFV, the two neutrinos can be of different flavor. Similar considerations apply for $O_{V_2}$. 
Under our simplifying assumptions, one can write Eq.\ (\ref{h-eff}) as
\be
{\cal H}_{\rm eff} = \frac{4G_F}{\sqrt{2}} V_{tb} V^*_{ts}C_{SM} 
\left[(1+C'_1) O_{V_1} + C'_2 O_{V_2}\right]
\ee
in terms of the scaled Wilson coefficients defined as 
$C'_{1,2} \equiv C_{V_{1,2}}/C_{SM}$, with 
\be
C_{SM} = \frac{\alpha}{2\pi\sin^2\theta_W} X_t(x_t)\,.
\ee
If the NP is also at the loop level, we expect $|C'_1|, |C'_2| \sim {\cal O}(1)$. If it is at tree level, $|C'_1|, |C'_2| \gg 1$. 
At the leading order, the Inami-Lim function $X_t$ is given by
\be
X_t^{LO} = \frac{x_t}{8} \left[ \frac{x_t+2}{x_t-1} -3 \frac{x_t-2}{(x_t-1)^2} \ln x_t\right]\,,
\ee
with $x_t = m_t^2/m_W^2$.

We will use the following numbers for our subsequent analysis:
\bea
&& m_B = 5.280~{\rm GeV}\,,\ \ m_{K^*}=0.896~{\rm GeV}\,,\nonumber\\
&& m_{K}=0.498~{\rm GeV}\,, \ \ 
m_s = 0.096~{\rm GeV} \,,\nonumber\\
&& |V_{tb}V_{ts}^*|=0.0401\,,\ \ \sin^2\theta_W = 0.2313\,,\nonumber\\
&& X_t=1.469\,,\ \ \tau_B = 1.519~{\rm ps}\,,\nonumber\\
&&G_F=1.166\times 10^{-5}~{\rm GeV}^{-2}\,,\ \ \alpha = 1/127.925\,,
\eea
where $\tau_B$ is the lifetime of the $B$ meson. 

\begin{widetext}

The exclusive differential decay distributions are given by \cite{buras-nu1,buras-nu2}
\bea
\frac{d\Gamma_{B \to K\nu\bar\nu}}{dq^2} &=& 
\frac{G_F^2 \alpha^2}{256\pi^5}\, \frac{|V_{tb} V_{ts}^*|^2 X_t^2}{m_B^3 \sin^4\theta_W} \, 
\lambda^{3/2}(m_B^2,m_K^2,q^2)
\left[f_+^K(q^2)\right]^2 \left\vert 1+ C'_1 + C'_2\right\vert^2\,,\nonumber\\
\frac{d\Gamma_{B \to K^*\nu\bar\nu}}{dq^2} &=& 
\frac{G_F^2 \alpha^2}{256\pi^5}\, \frac{|V_{tb} V_{ts}^*|^2 X_t^2}{m_B^3 \sin^4\theta_W} \, 
q^2\lambda^{1/2}(m_B^2,m_{K^*}^2,q^2)\times
\nonumber\\
&& \left[ \left( 
|1+C'_1|^2+|C'_2|^2 \right) \left(
H^2_{V,+} + H^2_{V,-} + H^2_{V,0} \right) 
-2 {\rm Re}[(1+C'_1){C'_2}^*] \left( H^2_{V,0} + 2 H_{V,+}H_{V,-} \right)
\right]\,,
\eea
and the inclusive distribution by 
\bea
\frac{d\Gamma_{B \to X_s\nu\bar\nu}}{dq^2} &=& \frac{G_F^2 \alpha^2}{128 \pi^5}\, 
 \frac{|V_{tb} V_{ts}^*|^2 X_t^2}{m_b^3 \sin^4\theta_W} \kappa(0) 
 \left(|1+C'_1|^2 + |C'_2|^2\right)\times\nonumber\\
 &&\lambda^{1/2}(m_b^2,m_s^2,q^2) \left[ 
 3q^2 \left(m_b^2+m_s^2 - q^2 -4m_sm_b\, \frac{ {\rm Re}[(1+C'_1){C'_2}^*]} 
 { |1+C'_1|^2 + |C'_2|^2} \right) + \lambda(m_b^2,m_s^2,q^2)\right]\,,
\eea 
\end{widetext}

where
\be
\lambda(a,b,c) = a^2+b^2+c^2-2\left( ab+bc+ca\right)\,,
\ee
and $\kappa(0) = 0.83$ is the QCD correction factor. Note that the structure of the interference term 
in $|1+C'_1|^2$ changes if $O_{V_1}$ has LFV or non-LFU nature.

  \begin{figure}[htbp]
\centering
  \includegraphics[scale=0.5]{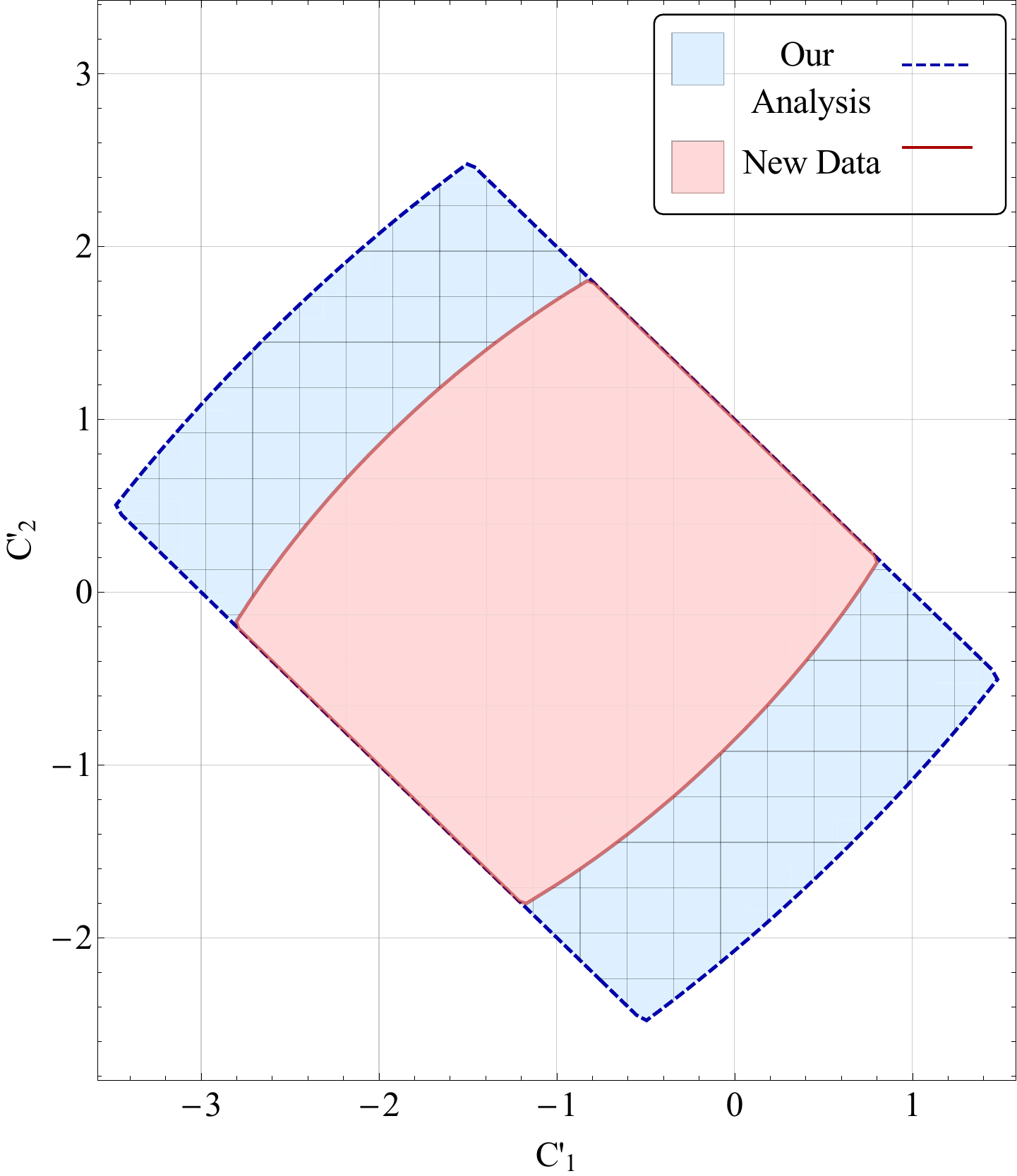} 
\caption{Allowed ranges of $C'_1$ and $C'_2$. Also shown is the truncated region allowed by the 
recent Belle data \cite{Grygier:2017tzo}.    }
\label{fig:c1c2}
\end{figure}

For $B\to K^*$, the form factors are defined in terms of the conventional set as
\begin{align}
H_{V,\pm}(q^2) &= 
M A_1(q^2) \mp \frac{\lambda^{1/2}(m_B^2, m_{K^*}^2,q^2) V(q^2)}{M}\,,\nn\\
H_{V,0}(q^2) &= \frac{M}{2m_{K^*}\sqrt{q^2}} \left[ 
\frac{ \lambda(m_B^2, m_{K^*}^2,q^2)}{M^2} A_2(q^2) \right. \nn \\
&\left.- \left(m_B^2 - m_{K^*}^2-q^2\right) A_1(q^2)\right]\,,
\end{align}
where $M = m_B+m_{K^*}$.
To get the form factors, one first defines the function
\be
z(q^2) = \frac{ \sqrt{t_+ - q^2} - \sqrt{t_+ - t_0}} 
{\sqrt{t_+ - q^2} + \sqrt{t_+ - t_0}}\,,
\ee
where
\be
t_{\pm} = \left(m_B \pm m_{K^{(*)}}\right)^2\,, \ \ 
t_0 = t_+ \left(1 - \sqrt{1-t_- / t_+}\right)\,.
\ee
Then we define the generic structure as 
\be
F_i(q^2) = \frac{1}{1-q^2/m_P^2} \sum_k \alpha^i_k \left[ z(q^2) - z(0)\right]^k\,,\label{fgen1}
\ee
where the pole masses $m_P$ are 
\be
V: 5.415~{\rm GeV},\
A_0: 5.366~{\rm GeV},\
A_1,A_{12}: 5.829~{\rm GeV},
\ee
and \cite{lcsr-aoife}
\begin{align}
\alpha_0^V &= 0.38(3),\ \  \alpha_1^V = -1.17(26),\ \ \alpha_2^V = 2.42\pm 1.53,\nonumber\\
\alpha_0^{A_0} &= 0.37(3),\  \alpha_1^{A_0} = -1.37(26),\ \alpha_2^{A_0} = 0.13\pm 1.63,\nonumber\\
\alpha_0^{A_1} &= 0.30(3),\  \alpha_1^{A_1} = 0.39(19),\ \ \ \alpha_2^{A_1} = 1.19\pm 1.03,\nonumber\\
\alpha_0^{A_{12}} &= 0.27(2),\alpha_1^{A_{12}} = 0.53(13),\ \ \alpha_2^{A_{12}} = 0.48\pm 0.66\,.
\label{ff2}
\end{align}

Here $A_2$ has been replaced by $A_{12}$, given by 
\begin{align}
&A_{12}(q^2) = \nn \\
&\frac{ 
M^2\left(m_B^2-m_{K^*}^2-q^2\right) A_1 - \lambda(m_B^2,m_{K^*}^2,q^2) A_2} 
{16 M m_B m_{K^*}^2}\,.
\end{align}
The form factor $A_0$ will be needed when we discuss the decays to light invisible scalars.

The scalar form factors $f_0$ and $f_+$ for $B \to K$ are given by \cite{Bouchard:2013pna}
\begin{align}
f_0(q^2) &= \sum_{k=0}^{K}a^0_k z(q^2)^k, \label{ffac1}\nonumber\\
f_+(q^2) &= \frac{1}{1-q^2/m_P^2} \times \nonumber\\
& \sum_{k=0}^{K-1}a^+_k \left[ z(q^2)^k-(-1)^{k-K}\frac{k}{K}z(q^2)^K\right] \,,
\end{align}
where
\be
m_P = m_B + \Delta^*_+
\ee
and
\bea
&& a_0^0=0.550(20),\ \ a_1^0=-1.89(23),\ \ a_2^0=1.98(1.24),\nonumber\\
&& a_3^0=-0.02(2.00),\ \ a_0^+=0.432(15),\ \ a_1^+=-0.65(23),\nonumber\\
&& a_2^+=-0.97(1.24),\ \ \Delta^*_+ =0.04578(35)\,.
\eea

Another observable that we may use is the modified 
transverse polarization fraction of $K^*$ in $B\to K^*\nu\bar\nu$ 
decays, defined as
\be
F'_T \equiv \frac{d\Gamma_T/dq^2}{\int (d\Gamma/dq^2)\, dq^2} = \tau_B \frac{d\Gamma_T}{dq^2}\,.
\ee
Note that the denominator has been integrated over, to give an 
overall normalization. It can be shown easily that
\begin{align}
\frac{d\Gamma_T}{dq^2}  &=  
\frac{G_F^2 \alpha^2}{256\pi^5}\, \frac{|V_{tb} V_{ts}^*|^2 X_t^2}{m_B^3 \sin^4\theta_W} \, 
q^2\lambda^{1/2}(m_B^2,m_{K^*}^2,q^2)\times \nn\\
& \left[ \left( 
|1+C'_1|^2+|C'_2|^2 \right) \left(
H^2_{V,+} + H^2_{V,-} \right) \right.\nn\\
&\left.-4 {\rm Re}[(1+C'_1){C'_2}^*] H_{V,+}H_{V,-}
\right]\,.
\end{align}

From the experimental bounds on the branching fractions, namely,
\begin{align}\label{resultOld}
\nn {\rm Br}(B\to K\nu\bar\nu) &< 1.7\times 10^{-5}\,,\ \\
{\rm Br}(B\to K^*\nu\bar\nu) &< 7.6\times 10^{-5}\,,
\end{align}
at 90\% CL, we get the following approximate 
constraints on the scaled Wilson coefficients, assuming them to be real (but not necessarily
positive):
\begin{align}
\nn -3.0 \, \leq C'_1+C'_2 &\leq \, 1.0\,,\ \\
\left( |1+{C'_1}|^2 + |C'_2|^2\right) -1.3 \left(1+C'_1\right) C'_2 \, &\leq \, 8.0\,.
\end{align}

The allowed parameter spaces are shown in Fig.\ \ref{fig:c1c2}.
Belle has a recent update \cite{Grygier:2017tzo}, mostly on the $B\to K^*\nu\bar\nu$ mode:
\begin{align}\label{resultNew}
\nn {\rm Br}(B\to K\nu\bar\nu) &< 1.6\times 10^{-5}\,,\ \\
{\rm Br}(B\to K^*\nu\bar\nu) &< 2.7\times 10^{-5}\,.
\end{align}
While our analysis uses the old parameter space, the results, as we will see, are obvious even with 
the new results. Fig.\ \ref{fig:c1c2} also shows the updated parameter space. 

We will show with the OO technique how much of the parameter space can be successfully 
differentiated from the SM, and with what confidence level. In our analysis, we have noted that 
the errors extracted on the new WCs are independent of the choices of the seed values. As mentioned 
earlier, these seed values can be chosen from the 
allowed NP parameter space. Obviously, depending on the data, different observables 
will have different power to differentiate NP effects from the SM. 
As the values are not known a priori, one has to look at all the observables and the pattern of the 
signal to have an idea of the underlying model.  

\subsection{Light invisible scalar}
\begin{figure*}[htbp]
\centering
  \subfloat[\label{scalpar05}]
  {\includegraphics[height=6cm]{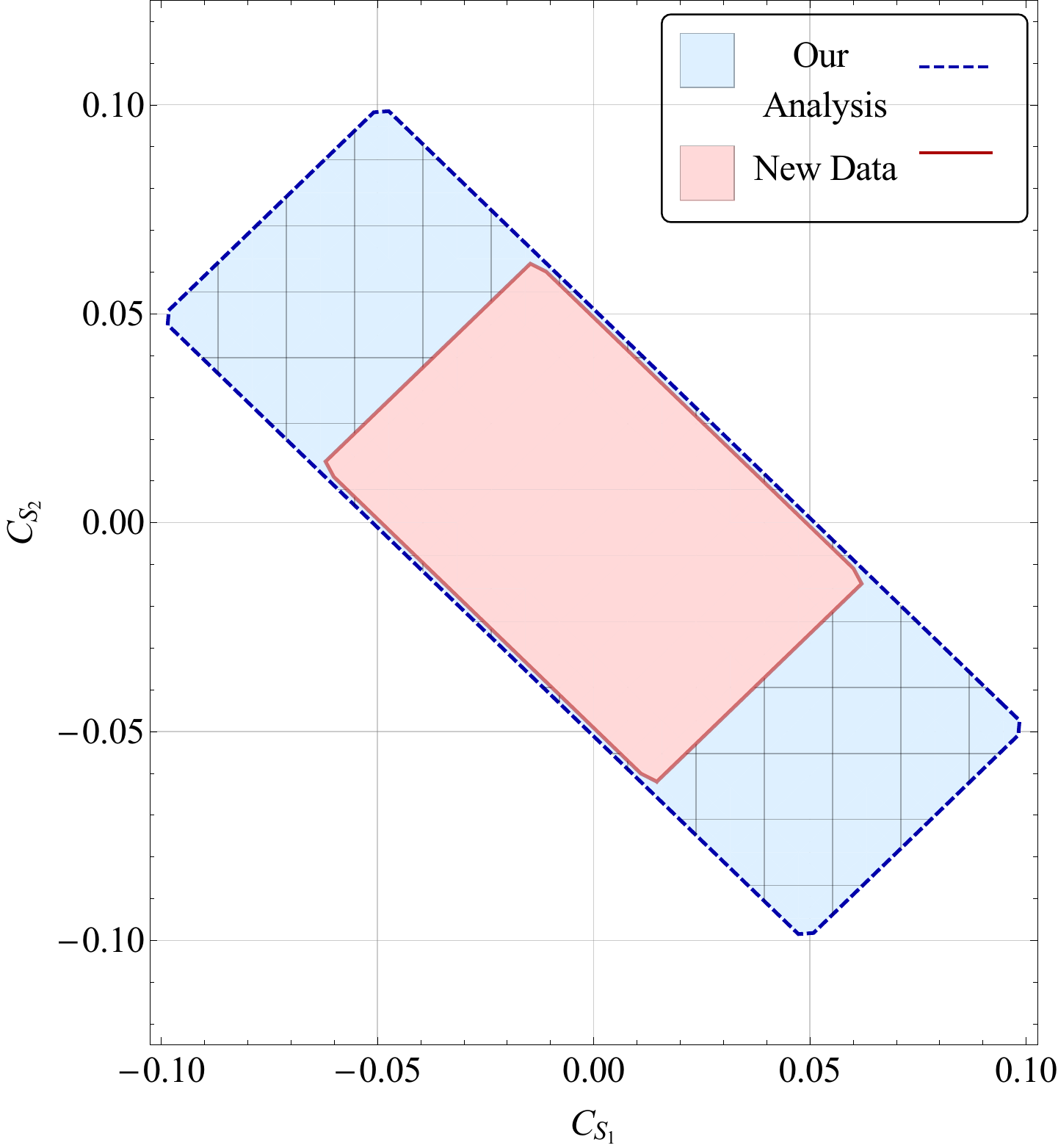}}
  \subfloat[\label{scalpar18}]
  {\includegraphics[height=6cm]{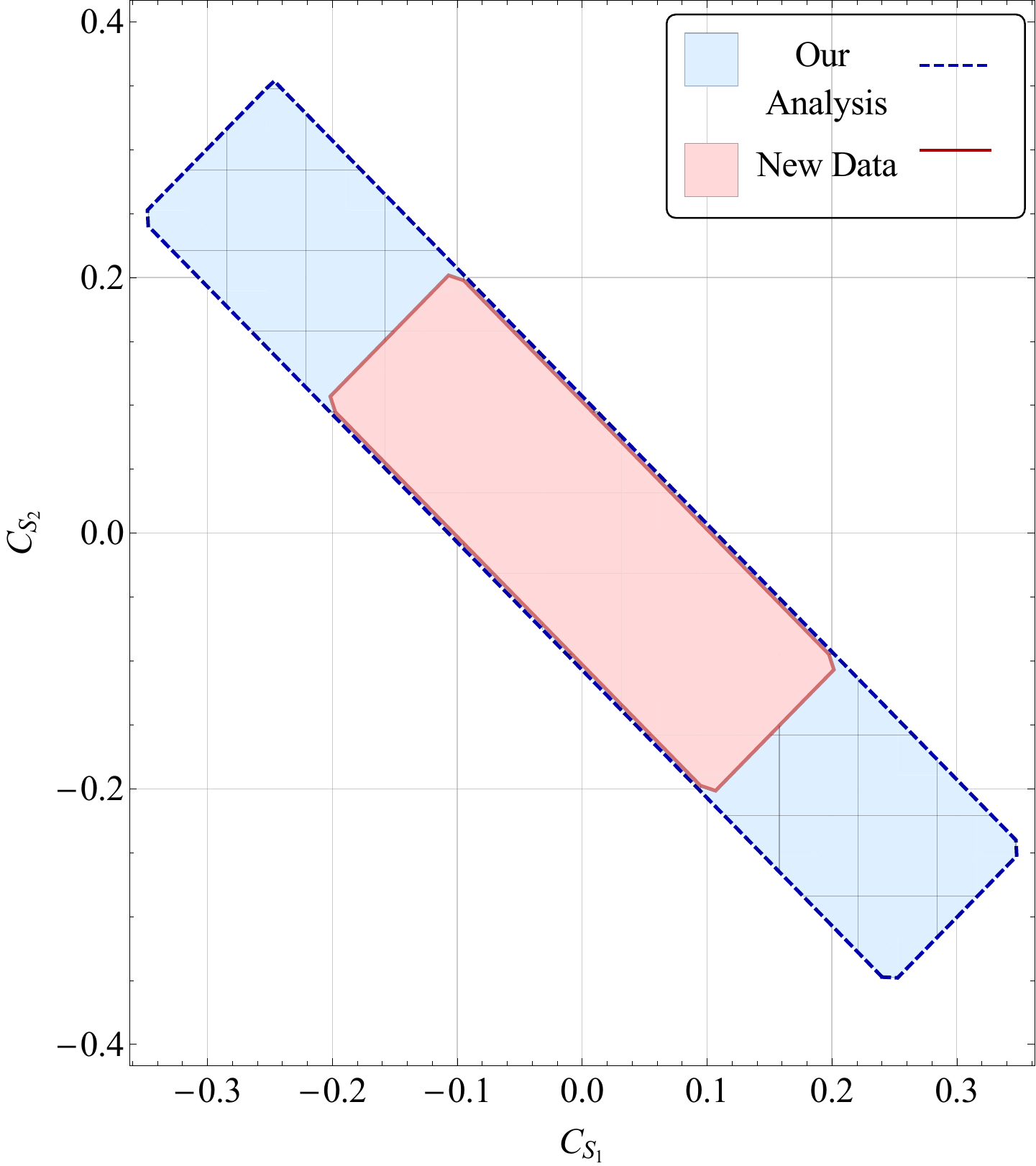}}
\caption{Allowed ranges of $C_{S_1}$ and $C_{S_2}$ for $m_S = 0.5$ GeV and $m_S=1.8$ GeV.
Also shown is the truncated region allowed by the recent Belle data \cite{Grygier:2017tzo}.  
}
\label{fig:scalarwc}
\end{figure*}


\begin{figure*}[htbp!]
  \centering
  \subfloat[\label{btkstnu} $B\to K^{\ast}\nu\bar\nu$ ]
  {\includegraphics[height=5.5cm]{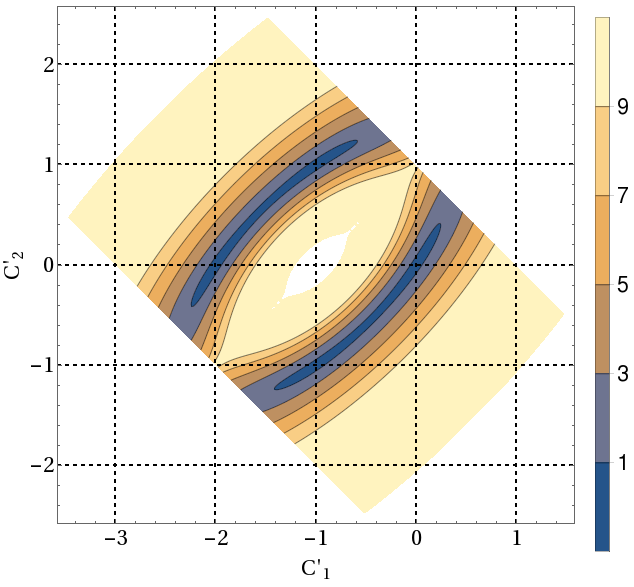}~
    \includegraphics[height=5.5cm]{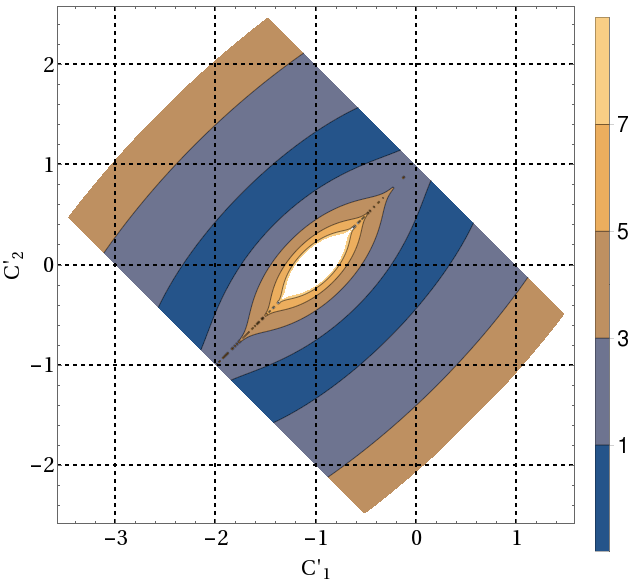}}\\
  \subfloat[\label{btxsnu} $B\to X_s\nu\bar\nu$ ]  
  {\includegraphics[height=5.5cm]{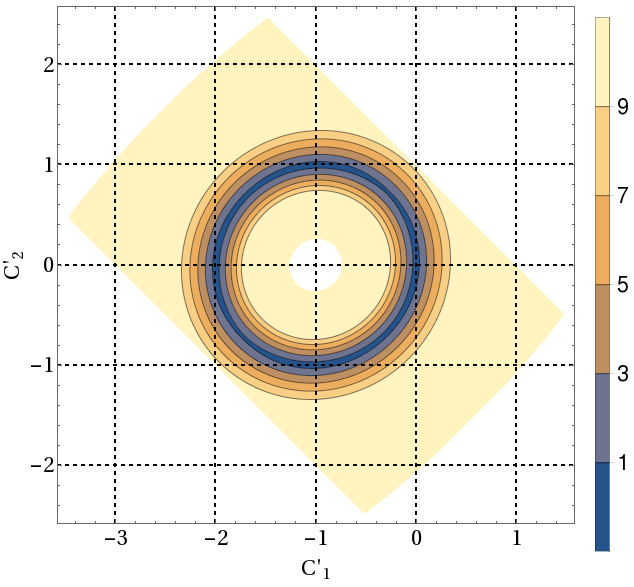}~
   \includegraphics[height=5.5cm]{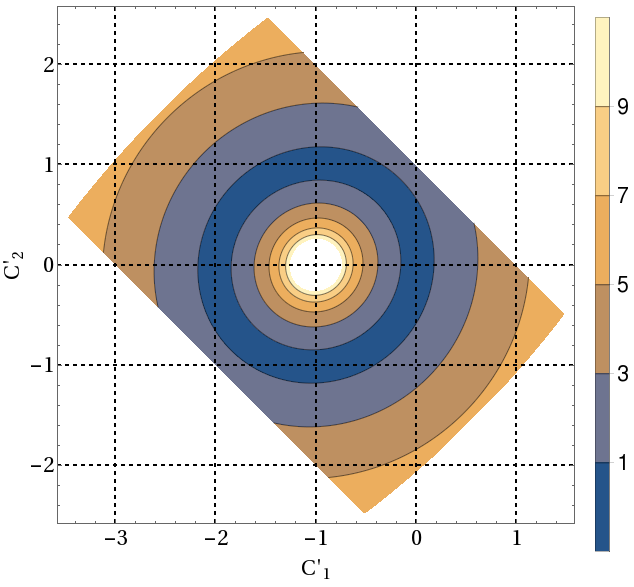}}\\
  \subfloat[ \label{drbtkstnu} Decay rate distributions: $B\to K^{\ast}\nu\bar\nu$ ]
  {\includegraphics[height=4.5cm]{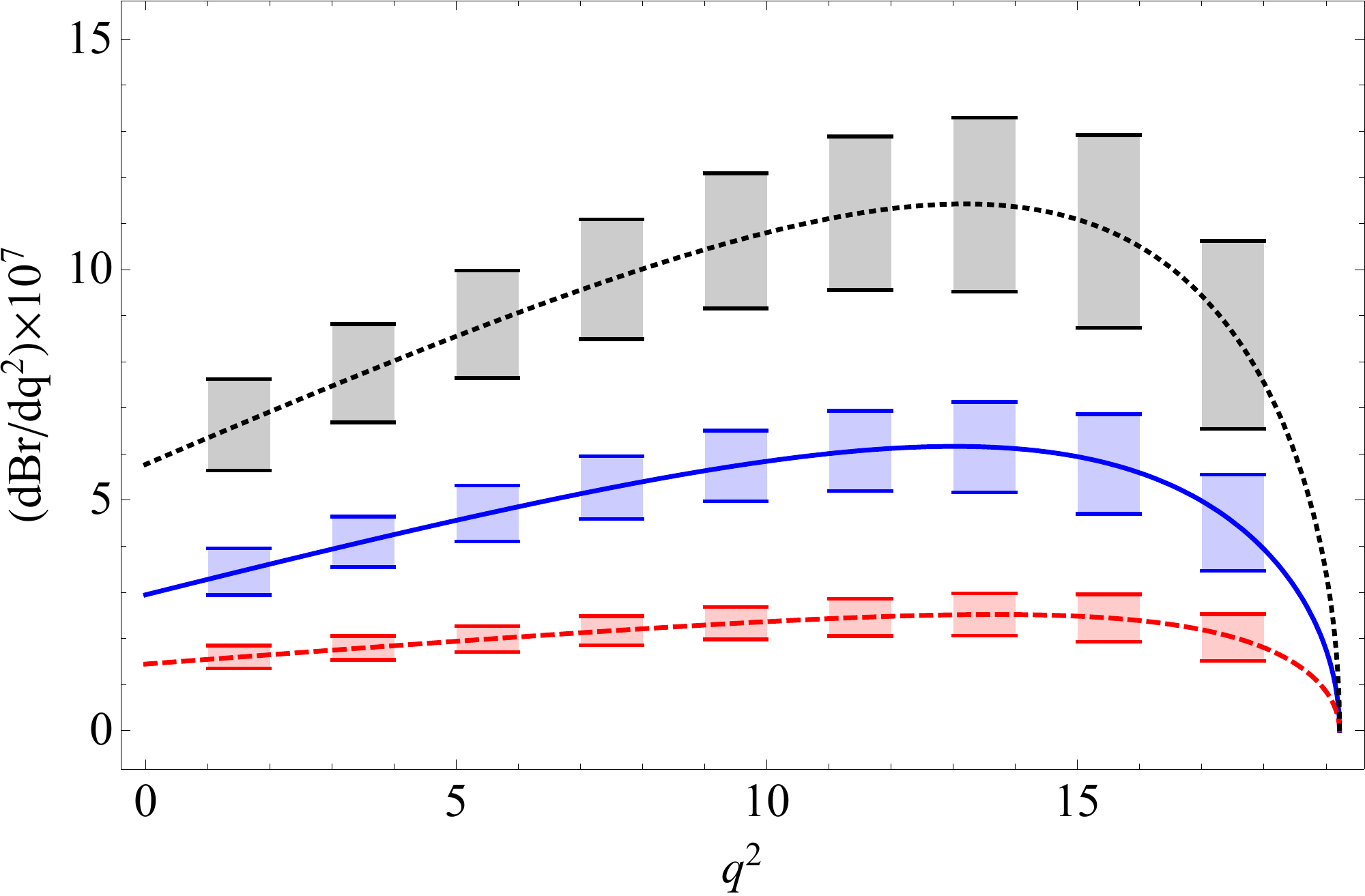}~}
  \subfloat[\label{drbtxsnu} Decay rate distributions: $B\to X_s\nu\bar\nu$ ]
   {\includegraphics[height=4.5cm]{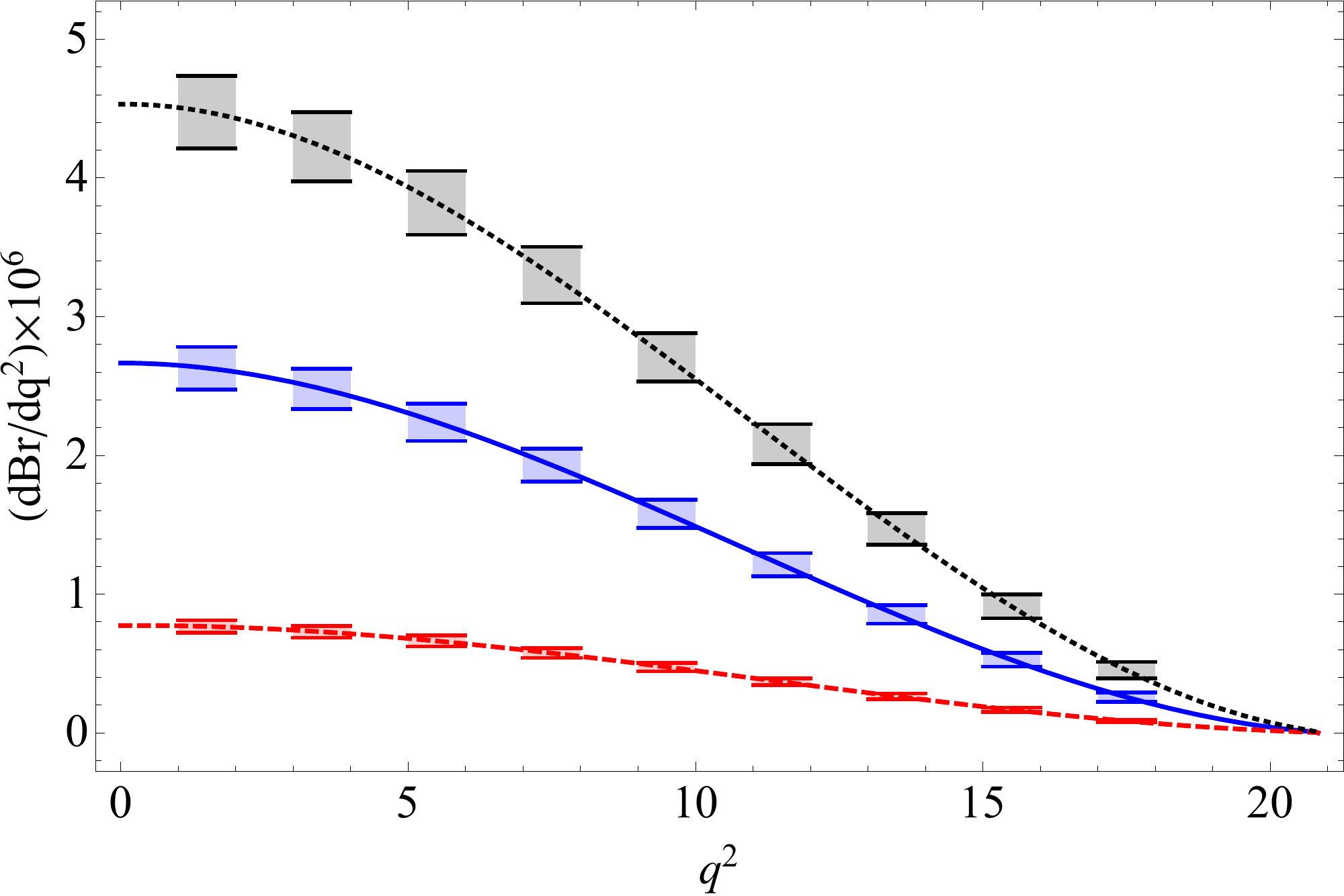}}~
      \includegraphics[height=4cm]{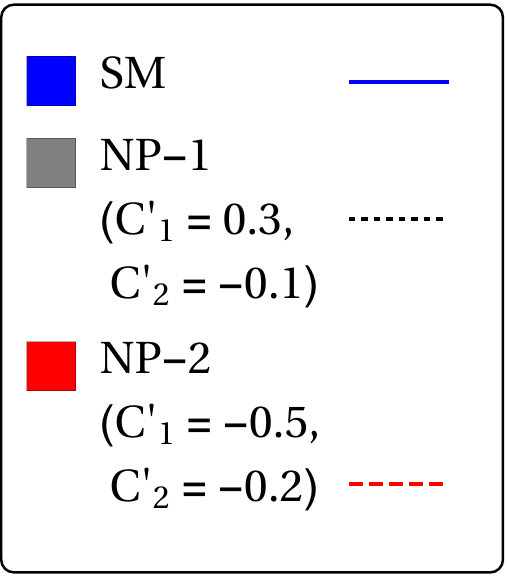}~
\caption{(a) and (b): The SM-NP differentiating $\chi^2$ contours for the exclusive and the inclusive 
channels coming from $b\to s \nu\bar\nu$, where the left and the right panels are for 
${\cal L}_{\rm int}= 50$ ab$^{-1}$ and 2 ab$^{-1}$ respectively. The $q^2$ (in GeV$^2$) distributions 
of the decay rates are shown in (c)  and (d) respectively, with 
${\cal L}_{\rm int}= 50$ for two benchmark scenarios of NP. For these and subsequent plots, we have 
not shown anything beyond $9\sigma$.}
\label{fig:bsnubnu}
\end{figure*}
 
\begin{figure*}[htbp]
\centering
 \includegraphics[height=5.5cm]{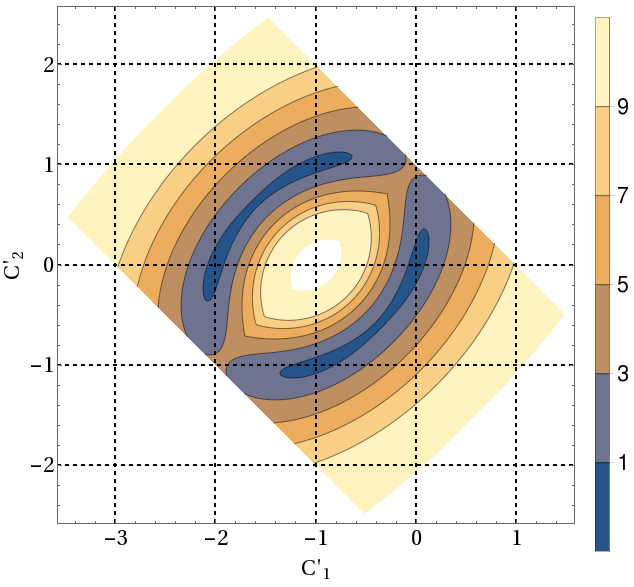}~
    \includegraphics[height=5.5cm]{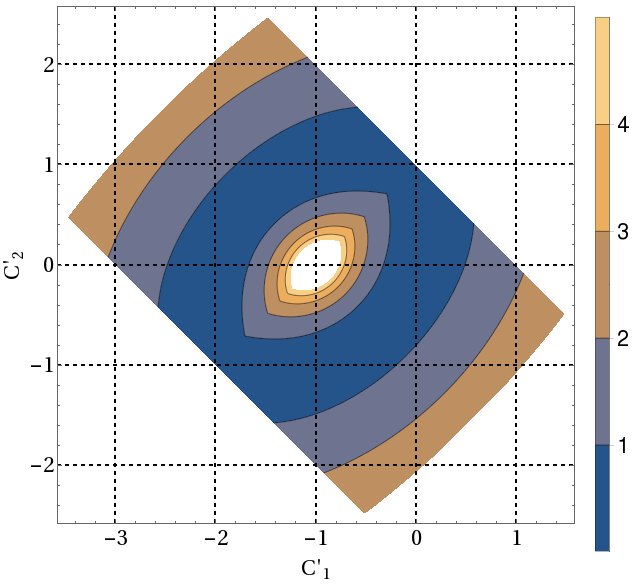}\\
  \caption{The contours from the measurement of $F^{\prime}_T$ for ${\cal L}_{\rm int}=50$ ab$^{-1}$ (left), 2 ab$^{-1}$ 
  (right). }
\label{fig:tranpolneu}
\end{figure*}

Another possibility is to consider the decay $b \to sSS$ where $S$ is some gauge singlet scalar, which can be 
a cold DM candidate. In the Higgs portal DM models, 
$S$ couples only to the SM doublet $\Phi$ through a term like $a_2 S^2\Phi^\dag\Phi 
\to \frac12 a_2 S^2 h^2$ in the Lagrangian, where $h$ is the SM Higgs field. If $m_S < m_h/2$, the invisible 
decay $h\to SS$ opens up, and one must keep $a_2$ to be sufficiently small to avoid the LHC bound on 
such invisible decay channels: ${\rm BR}(h\to {\rm invisible}) < 10\%$. Although this is in contradiction to 
a thermalized cold DM giving the correct relic density of the universe, the singlets can form only a part of the 
relic density and may even be non-thermal in nature. 

One has to be careful about the construction of effective operators. At the first sight, it may appear that
an effective dimension-6 operator $\bar{s}_L b_R \Phi S^2$ may lead to the decay $b\to sSS$ when 
$\Phi$ is replaced by its vacuum expectation value (VEV). This is indeed the case if $S$ does not have 
any VEV, which is essential if $S$ is a DM candidate (otherwise it will mix with $h$ and decay to SM 
final states). On the other hand, the Higgs penguin diagrams like $b\to sh^*$, $h^*\to SS$, as discussed 
in some literature \cite{willey,bird}, cannot be there if the electroweak symmetry is broken by a single 
Higgs field. The reason is that the effective off-diagonal Yukawa coupling $y_{bsh}$ is proportional to
the off-diagonal mass term $m_{bs}$ in the mass matrix, and once one goes to the stationary basis, 
such off-diagonal Yukawa couplings must vanish. This loophole can be avoided if there are more than one
fields responsible for symmetry breaking, or if there are higher dimensional operators involving $\Phi$ 
in quadratic or more, so that the proportionality of the Yukawa matrix and the mass matrix gets spoiled
\footnote{An example is provided in Ref.\ \cite{hmutau}.}. 
Here, we will just assume the existence of a set of effective operators and explore the consequences. 
 
We start with an effective Lagrangian of the form  
\be
{\cal L}_{b\to sSS} = C_{S_1} m_b \bar{s}_L b_R S^2 + C_{S_2} m_b \bar{b}_L s_R S^2 + {\rm H.c.}\,,
\ee
and assume only the SM operator to be present for the $b\to s\nu\bar\nu$ decay, so that
\be
\left. \frac{d\Gamma}{dq^2} \right\vert_{b\to s+{\rm invis}} 
= \left. \frac{d\Gamma}{dq^2} \right\vert_{b\to s\nu\bar\nu} +  
\left. \frac{d\Gamma}{dq^2} \right\vert_{b\to sSS}\,.
\ee
\begin{widetext}
Following Ref.\ \cite{buras-nu1}, one gets 
\bea
\frac{d\Gamma_{B \to KSS}}{dq^2} &=& 
\frac{ f_0^2(q^2) (m_B^2-m_K^2)^2 \left\vert C_{S_1}+C_{S_2}\right\vert^2}
{512\pi^3 m_B^3} \sqrt{1-\frac{4m_S^2}{q^2}}  \lambda^{1/2}(m_B^2,m_{K}^2,q^2)\,,\nonumber\\
\frac{d^2\Gamma_{B \to K^*SS}}{dq^2\, d\cos\theta} &=& 
\frac{ 3 A_0^2(q^2) \left\vert C_{S_1}-C_{S_2}\right\vert^2}
{1024\pi^3 m_B^3} \sqrt{1-\frac{4m_S^2}{q^2}}  \lambda^{3/2}(m_B^2,m_{K^*}^2,q^2) \, \cos^2\theta\,,\nonumber\\
\frac{d\Gamma_{B \to X_sSS}}{dq^2} &=& \frac{|C_{S_1}|^2 + |C_{S_2}|^2} {128 \pi^3 m_b}\,
\sqrt{1-\frac{4m_S^2}{q^2}} \lambda^{1/2}(m_b^2,m_s^2,q^2) \left[ (m_b^2+m_s^2-q^2) - 4 m_s m_b \frac{ {\rm Re}[C_{S_1}C_{S_2}^*]} 
 { |C_{S_1}|^2 + |C_{S_2}|^2}\right]\,,
\label{bssincl}
\eea
where the form factors $f_0(q^2)$
and $A_0(q^2)$ can be obtained from Eq.\ (\ref{ffac1}) and Eq.\ (\ref{fgen1}) respectively.

For the decay $B\to K^* SS$, all $K^*$s are longitudinally polarized. We define a modified longitudinal 
polarization fraction
\be
F'_L \equiv \frac{d\Gamma_L/dq^2}{\int(d\Gamma/dq^2)\,dq^2} = \tau_B \frac{d\Gamma_L}{dq^2}
=\tau_B \left[ \frac{d\Gamma_L}{dq^2}_{B\to K^*\nu\bar\nu} +  \frac{d\Gamma}{dq^2}_{B\to K^*SS}
\right]\,,
\label{flmod}
\ee
which comes out to be 
\begin{align}
F'_L  &=  
\frac{G_F^2 \alpha^2}{256\pi^5}\, \frac{|V_{tb} V_{ts}^*|^2 X_t^2}{m_B^3 \sin^4\theta_W} \tau_B  \, 
q^2\lambda^{1/2}(m_B^2,m_{K^*}^2,q^2) \, H_{V,0}^2 
\nonumber\\
& + \left\vert C_{S_1}-C_{S_2}\right\vert^2 \frac{A_0^2(q^2)}{512 \pi^3 m_B^3} \tau_B 
\, \sqrt{1-
\frac{4m_S^2}{q^2} }\,  \lambda^{3/2}(m_B^2,m_{K^*}^2,q^2)\,.
\end{align}

Obviously, the allowed range of the WCs depend on the scalar mass $m_S$, which is shown in 
Fig.\ \ref{fig:scalarwc}. Thus, apart from the new Wilson coefficients, $m_S$ is also another {\em a priori} 
unknown quantity. 
 \end{widetext}

\begin{figure*}[htbp]
\centering
\subfloat[\label{bkss} $B\to K SS$ ]
 { \includegraphics[height=5.5cm]{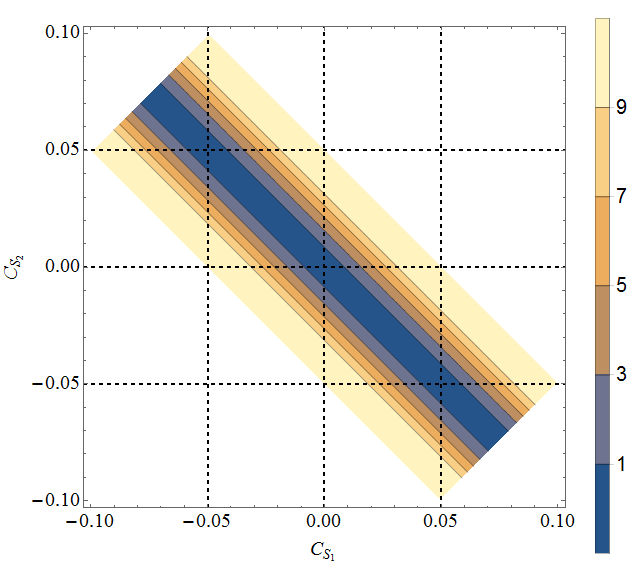}~
    \includegraphics[height=5.5cm]{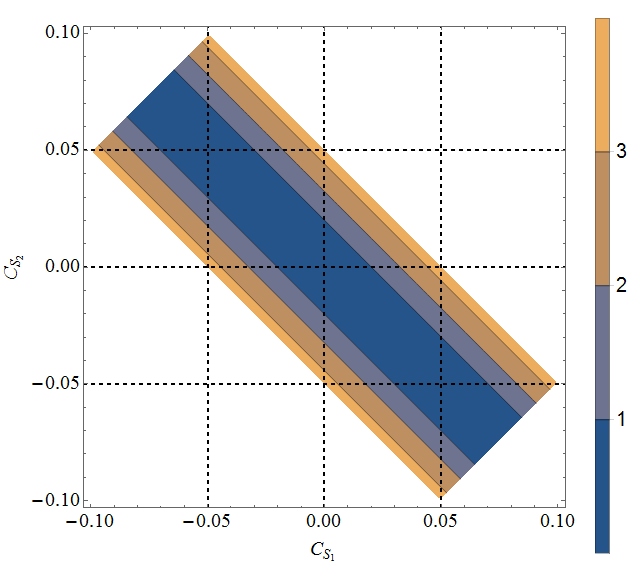}}\\
 \subfloat[\label{bkstss} $B\to K^{\ast} SS$ ] 
  {\includegraphics[height=5.5cm]{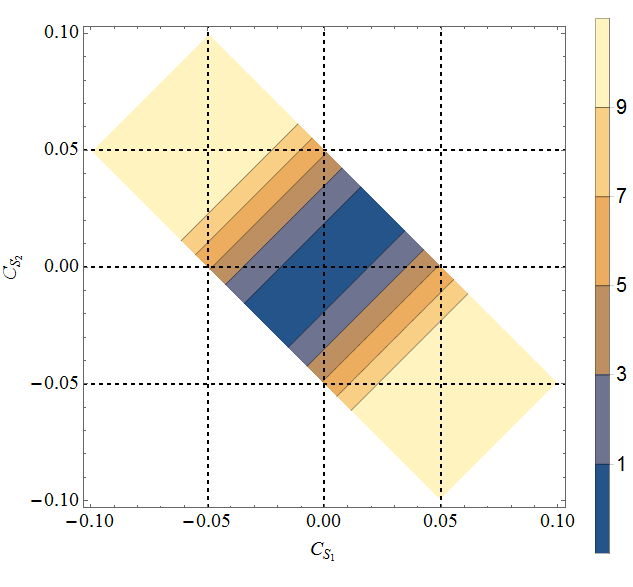}~
    \includegraphics[height=5.5cm]{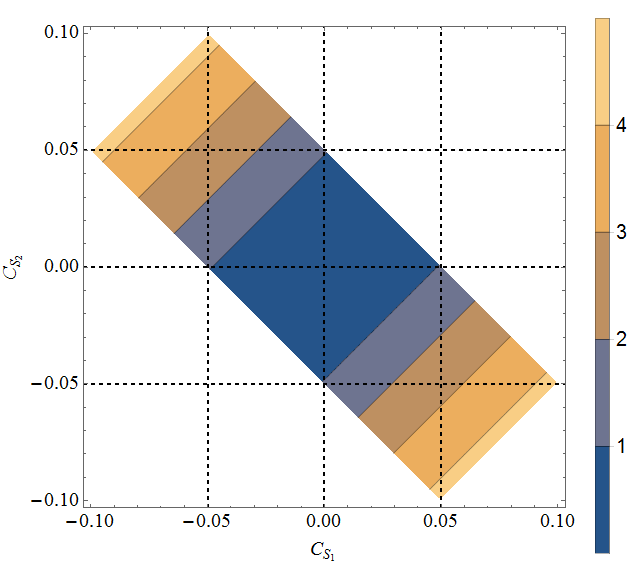}}\\
   \subfloat[\label{bxsss} $B\to X_s SS$ ]
  {\includegraphics[height=5.5cm]{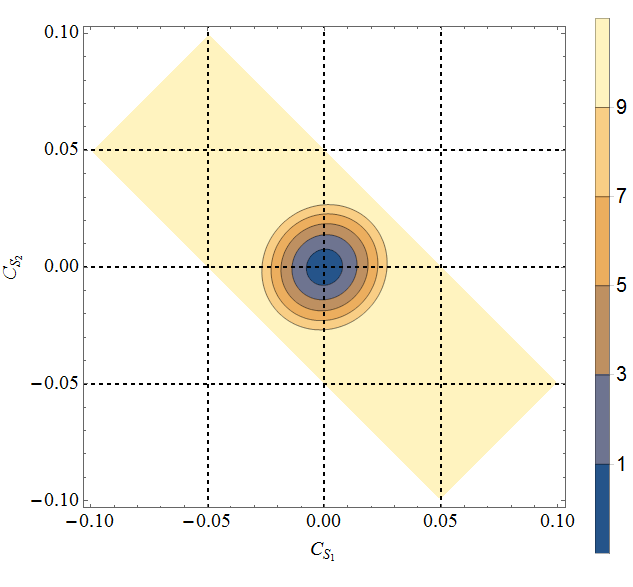}~ 
    \includegraphics[height=5.5cm]{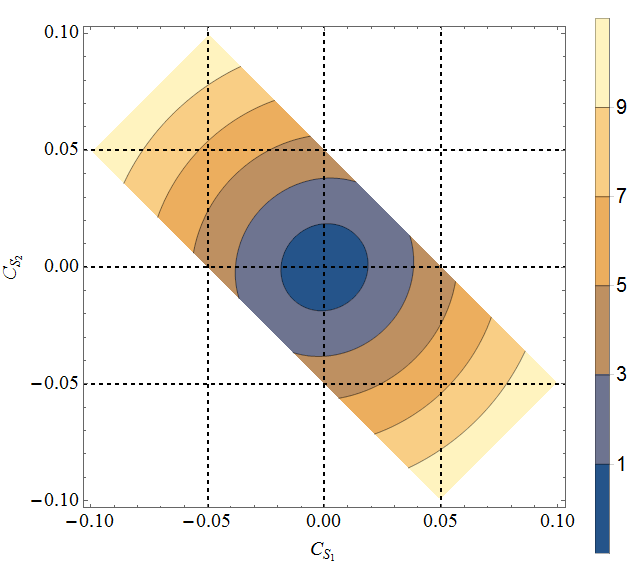}} \\
\caption{The differentiation contours for $b\to sSS$ with $m_S=0.5$ GeV and 
${\cal L}_{\rm int}=50$ ab$^{-1}$ (left panels),
2 ab$^{-1}$ (right panels). }
\label{fig:bsSS05}
\end{figure*}

\section{Results: Only neutrinos}

Our results are shown for the projected SuperBelle integrated luminosity ${\cal L}_{\rm int} = 
50$ ab$^{-1}$. However, 
to motivate experimentalists, we also show, for some cases, the results with ${\cal L}_{\rm int} = 2$
ab$^{-1}$, just to bring home the message that there might be reasons to feel excited even within
the first year of running. 
We have taken the production cross-section for $B^0$ and $B^+$ to be the same, which is known to 
be an excellent approximation. 
 The detection
efficiencies for different channels are taken from Ref.\ \cite{belle-knunu} \footnote{These are for Belle-I. 
The efficiencies are expected to go up for Belle-II, but we have been conservative in our estimates.}:
\bea
\epsilon(B^+\to K^+\nu\bar\nu) &=& 5.68\times 10^{-4}\,,\nonumber\\
\epsilon(B^0\to K_S\nu\bar\nu) &=& 0.84\times 10^{-4}\,,\nonumber\\
\epsilon(B\to K^\ast\nu\bar\nu) &=& 1.46\times 10^{-4}\,,
\eea
and we use the $SU(2)$ averaged detection efficiency for 
$B\to K^*\nu\bar\nu$. We also take the detection efficiency 
for the semi-inclusive $B\to X_s$ channel to be the same as that of $B\to K^*$. These numbers
will probably be slightly modified for the next generation detectors. However, in the absence of a detailed 
simulation study, it is impossible to include the systematic errors, so we have to work with the 
statistical error only.

\begin{figure*}[htbp]
\centering
\subfloat[\label{drbkss} $B\to K SS$ ]
{\includegraphics[height=5cm]{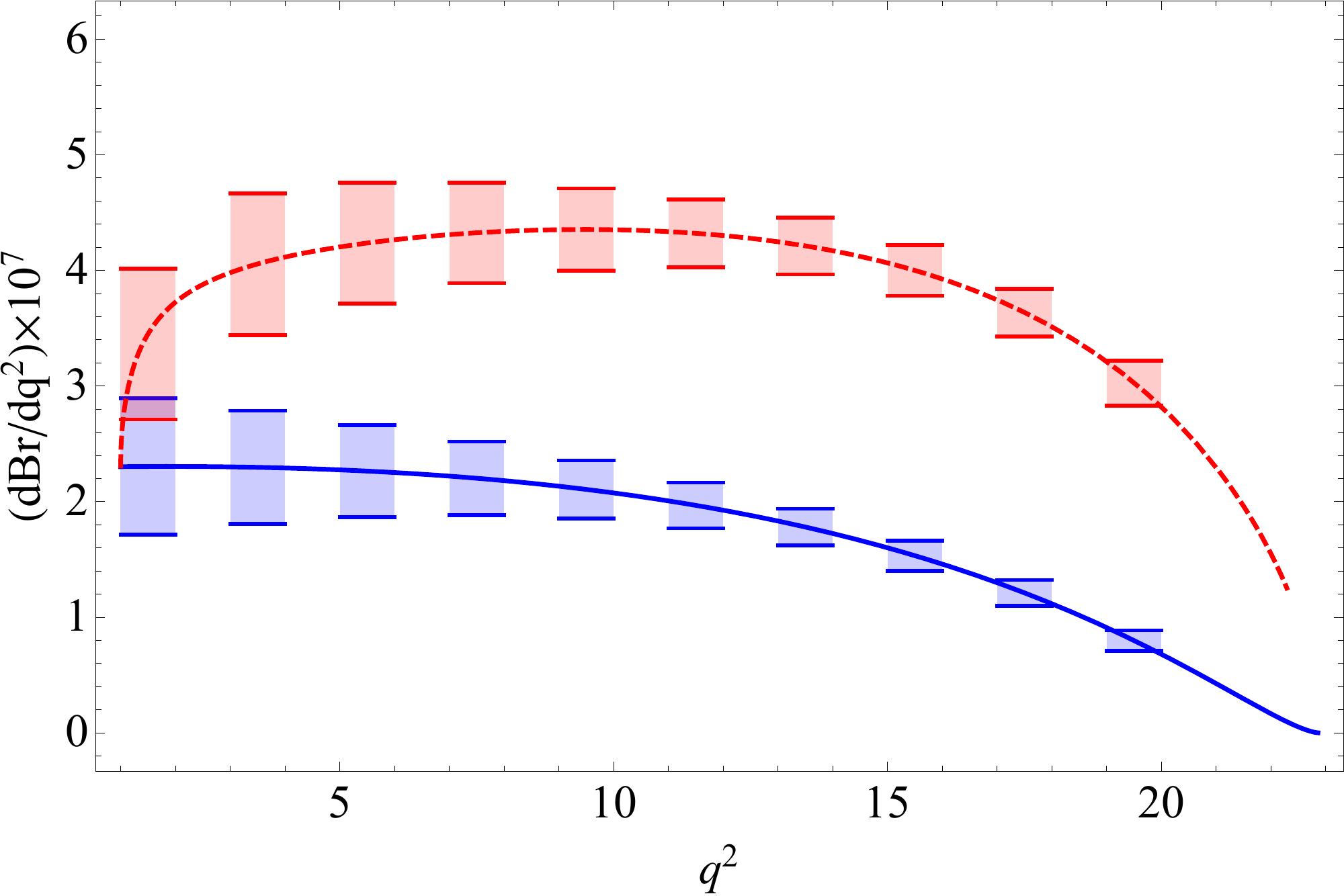}}~
\subfloat[\label{drbkstss} $B\to K^{\ast} SS$ ]
 {\includegraphics[height=5cm]{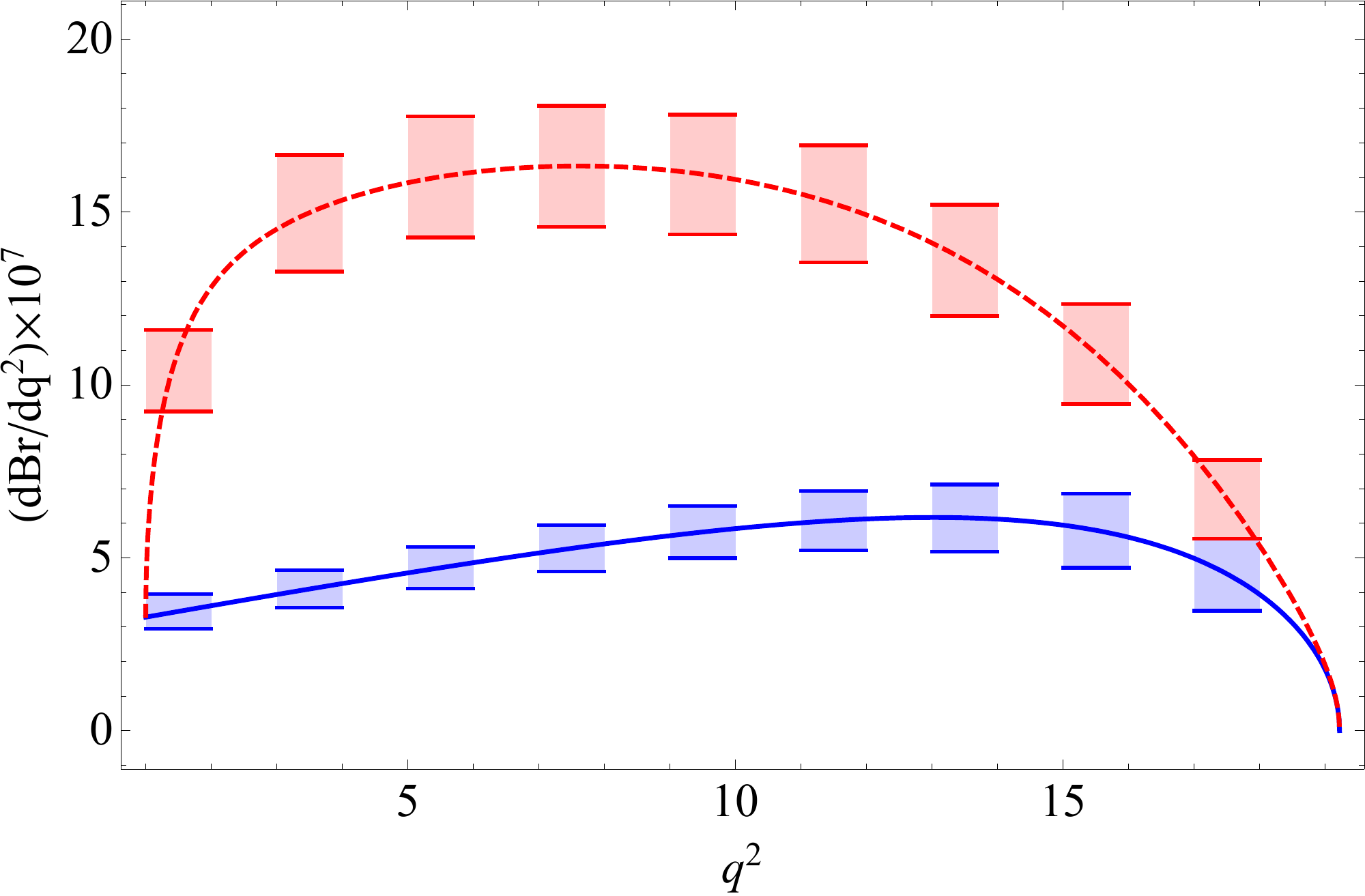}}\\
 \subfloat[\label{drbxsss} $B\to X_s SS$ ]
{\includegraphics[height=5cm]{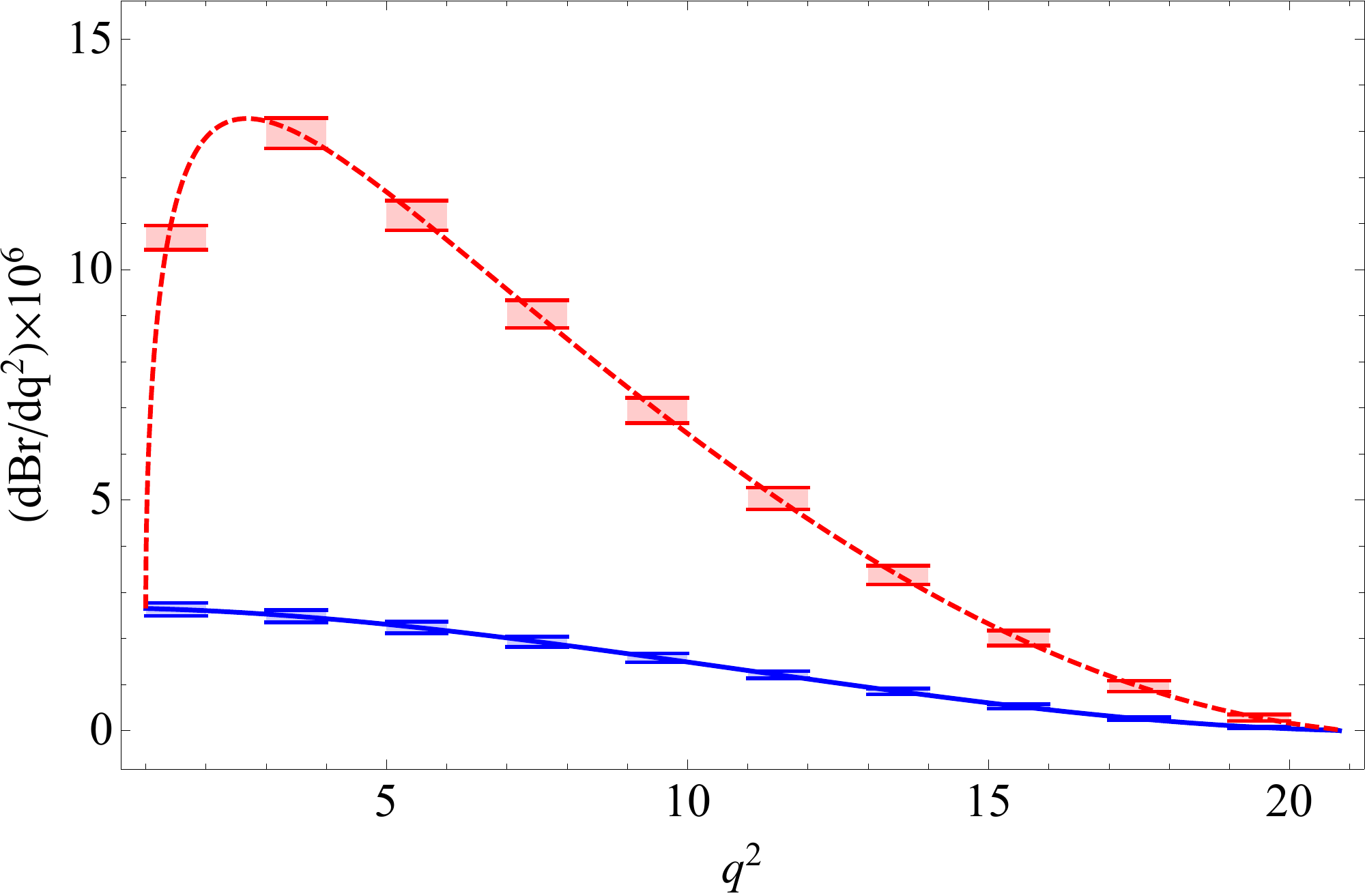}}~
 \includegraphics[height=3cm]{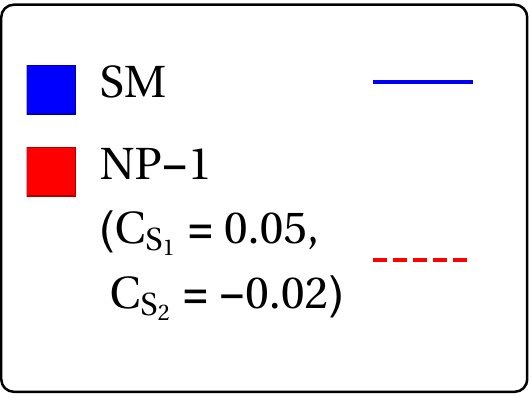}~
\caption{Comparison of the $q^2$ (in GeV$^2$) distributions of the decay rates for several $b\to sSS$
channels, with ${\cal L}_{\rm int}=50$ ab$^{-1}$ and $m_S = 0.5$ GeV.}
\label{fig:drbsSS05}
\end{figure*}
 
\begin{figure*}[htbp]
\centering
  \subfloat[ $B\to K SS$ ]
 { \includegraphics[height=5.5cm]{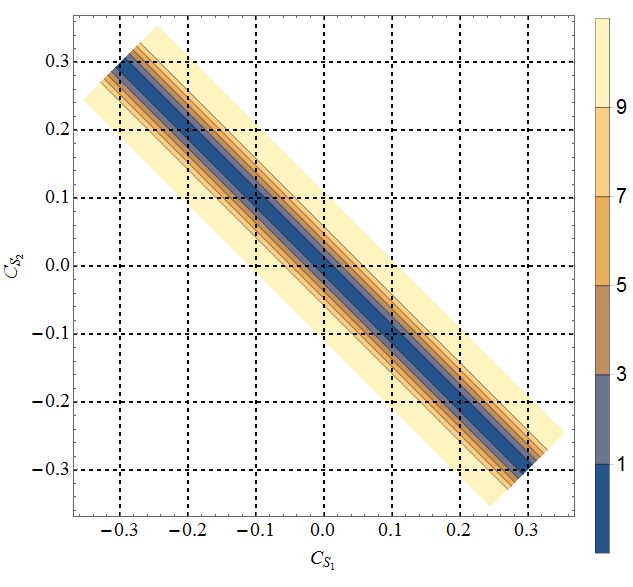}~
    \includegraphics[height=5.5cm]{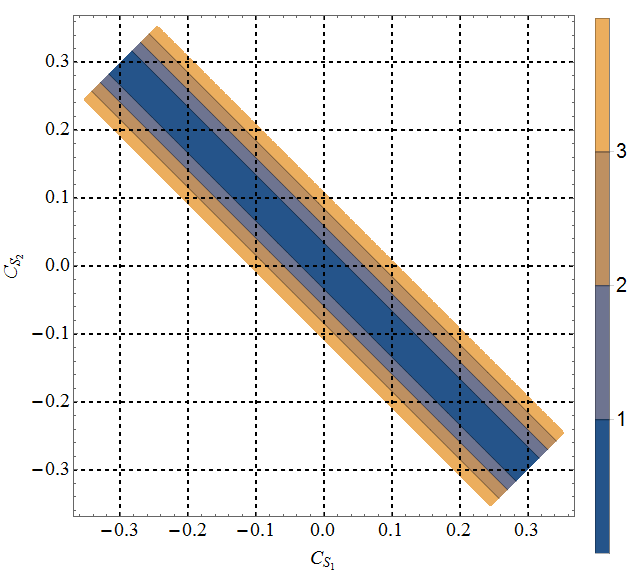}}\\
 \subfloat[ $B\to K^{\ast} SS$ ] 
  {\includegraphics[height=5.5cm]{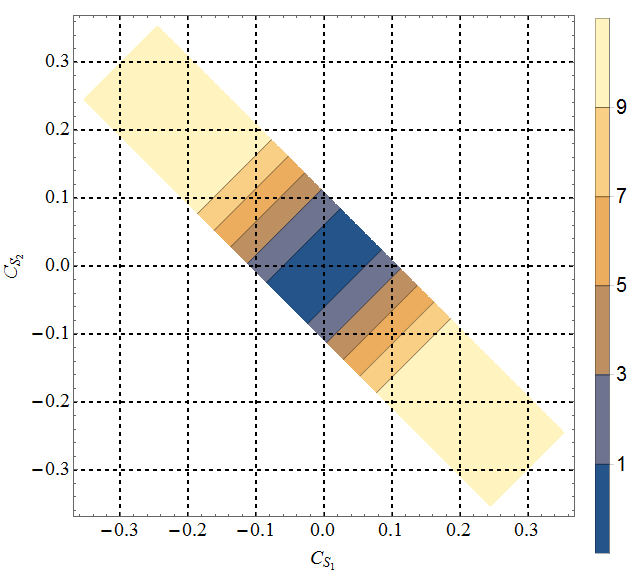}~
    \includegraphics[height=5.5cm]{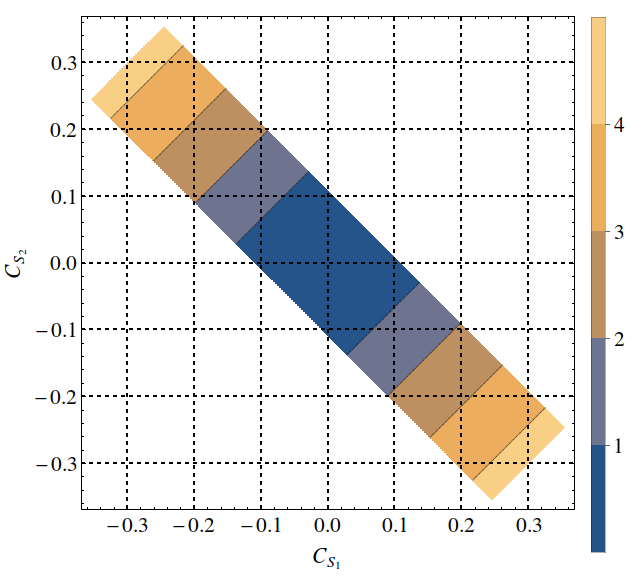}}\\
   \subfloat[ $B\to X_s SS$ ]
  {\includegraphics[height=5.5cm]{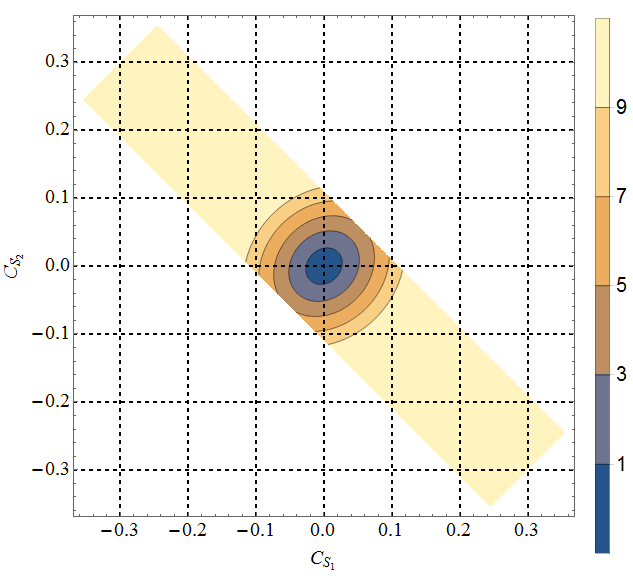}~ 
    \includegraphics[height=5.5cm]{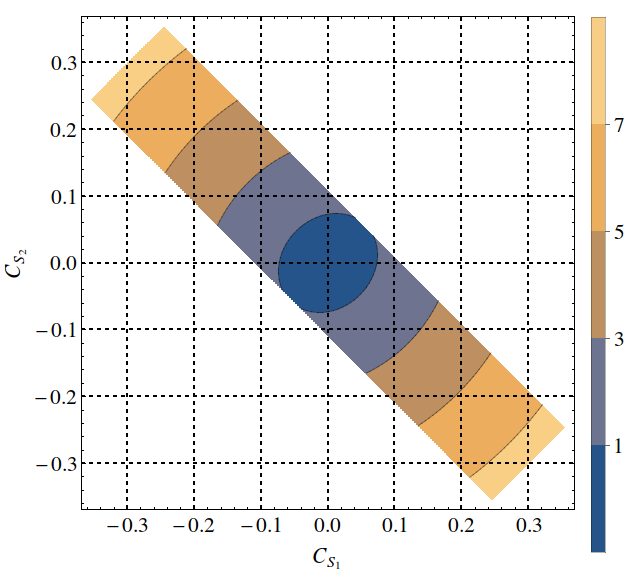}} \\
\caption{Same as Fig.\ \ref{fig:bsSS05} with $m_S=1.8$ GeV. }
\label{fig:bsSS1.8}
\end{figure*}

 \begin{figure*}[htbp]
\centering
\subfloat[ $B\to K SS$ ]
{\includegraphics[height=5cm]{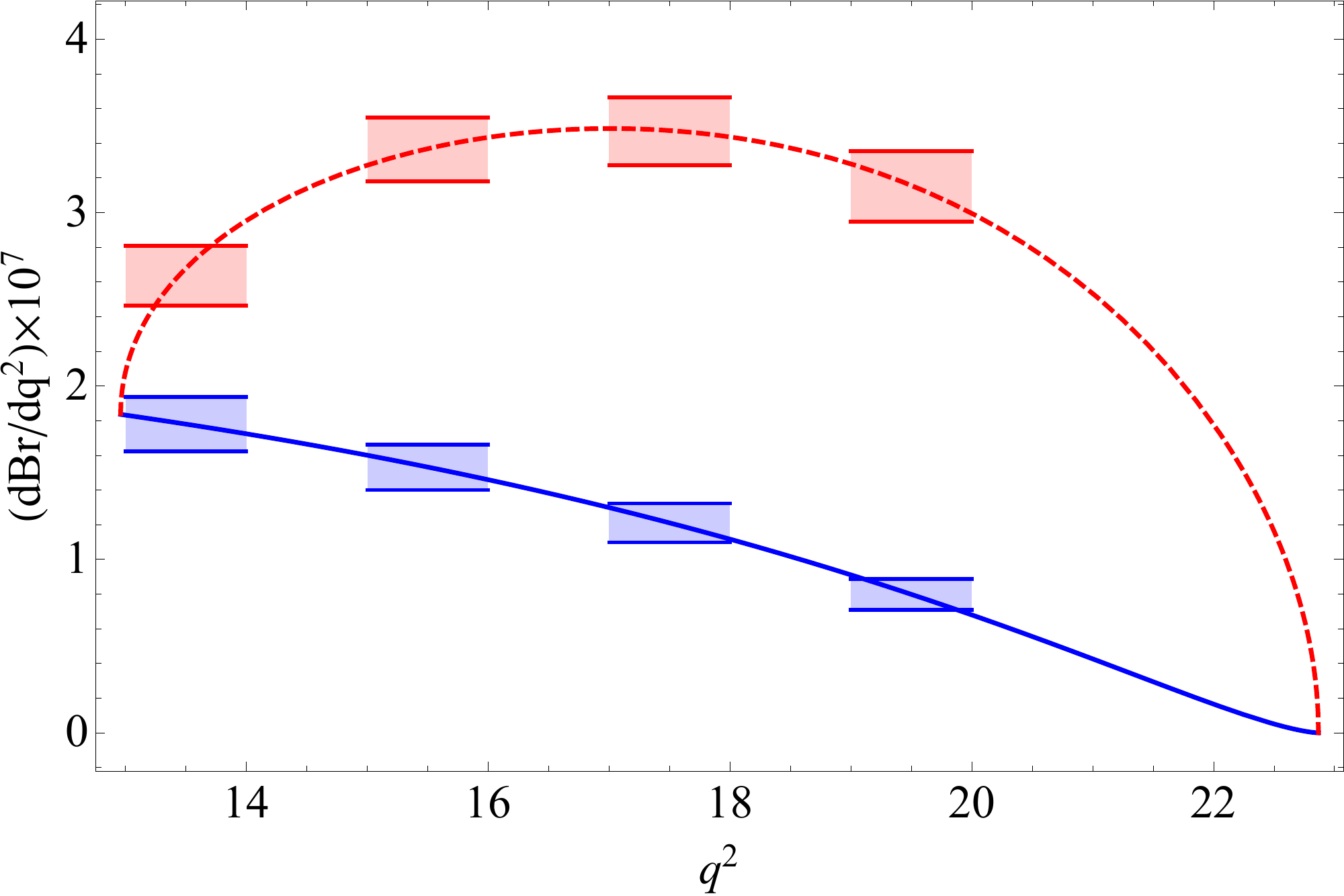}}~
\subfloat[ $B\to K^{\ast} SS$ ]
 {\includegraphics[height=5cm]{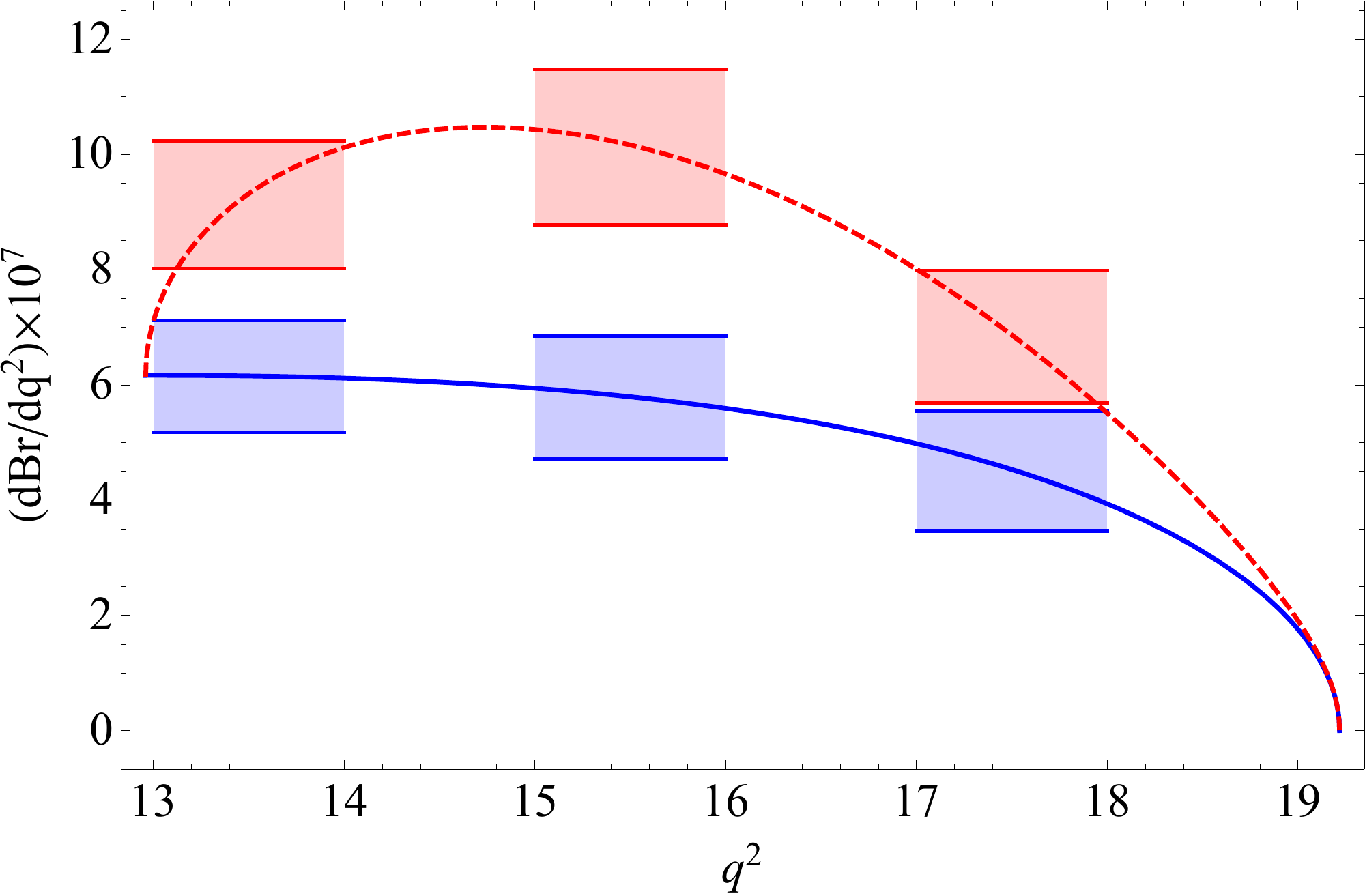}}\\
 \subfloat[ $B\to X_s SS$ ]
{\includegraphics[height=5cm]{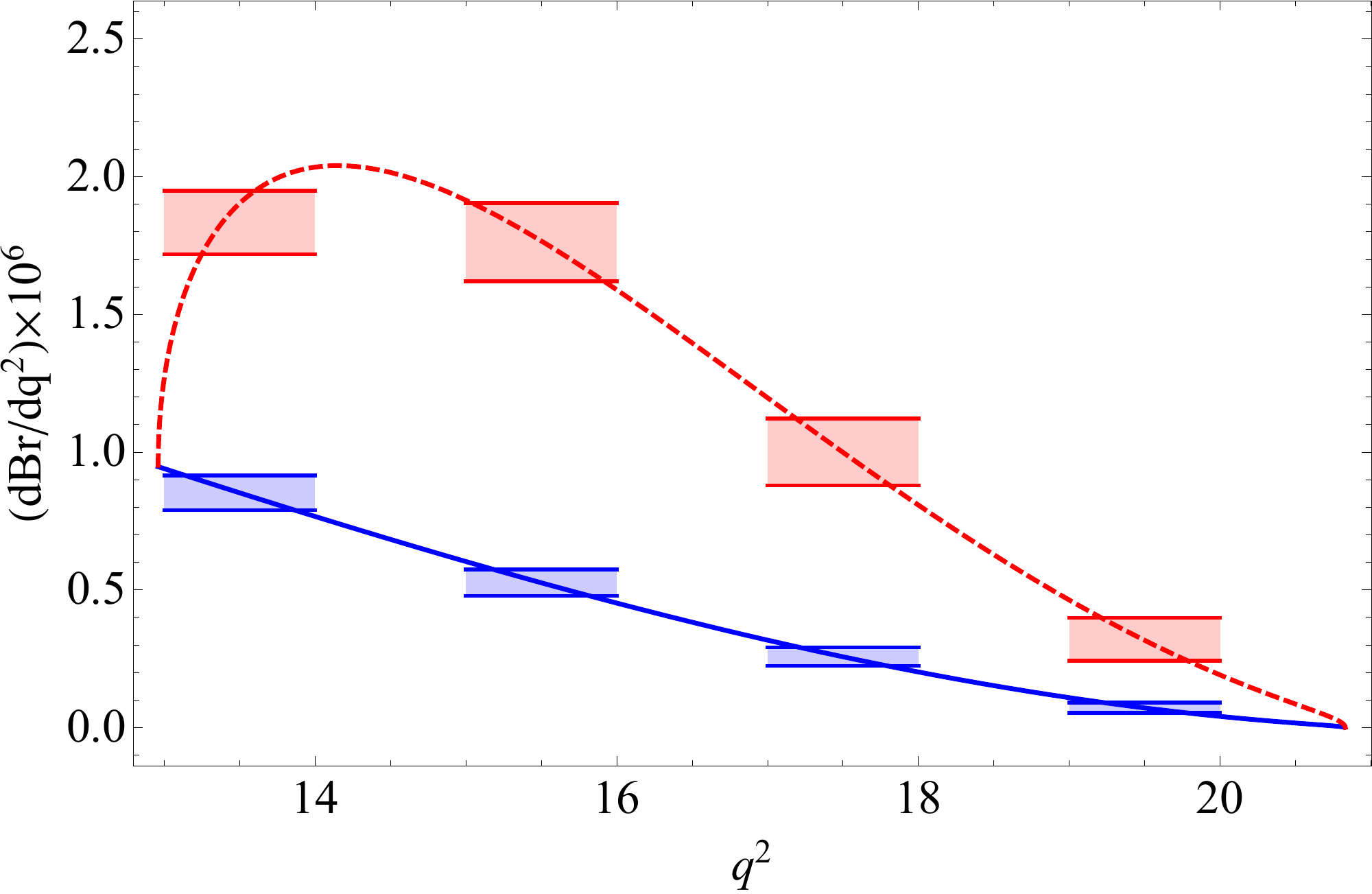}}~
 \includegraphics[height=3cm]{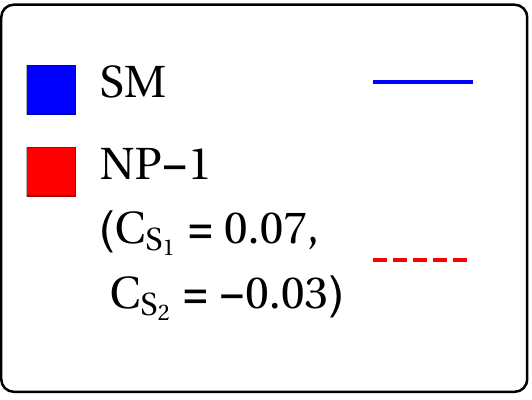}~
\caption{Same as Fig. \ref{fig:drbsSS05} with $m_S=1.8$ GeV.}
\label{fig:drbsSS1.8}
\end{figure*}

From the definition of the OOs, it is clear that one needs to have at least two different $c_i$s for this
technique to work, otherwise it is just a simple scaling of the SM expectation. With the assumption of 
LFU and no LFV, this is what happens for the decay $B\to K\nu\bar\nu$; the SM factor of 1 is replaced by
$|1+C'_1+C'_2|^2$. On the other hand, the decays $B\to K^\ast\nu\bar\nu$, $B\to X_s\nu\bar\nu$, and the 
scaled transverse polarization fraction $F'_T$ all have more than one combinations of the new WCs.

In Figs.\ \ref{btkstnu} and \ref{btxsnu}, we show the results from the OO analysis of the decays 
$B\to K^\ast\nu\bar\nu$ and $B\to X_s\nu\bar\nu$ respectively. 
These plots use the $q^2$-integrated data, and shows how far the NP can be 
differentiated from the SM depending on the precise values of $C'_1$ and $C'_2$. 
The $\chi^2 = n^2$ (with n = 1,3,5,7,9 ) 
lines are obtained in the $C_1^{\prime}, C_2^{\prime}$ basis with $\chi^2|_{SM} = 0$, where, depending 
on the values of $n$, each line represents a deviation of $n~\sigma$ from the SM. 
Obviously, $C'_1=C'_2=0$ is the SM and close to that 
the chances of separation are the weakest, as shown by the $1\sigma$ band. 
Note that $C'_1=-2$ and $C'_2=0$ is also 
SM-like, because of the destructive interference between the two amplitudes, keeping $|1+C'_1|^2 = 1$. 
As can be seen, with ${\cal L}_{\rm int}=50$ ab$^{-1}$ (left panels), both the decays can differentiate 
NP from SM over most of the allowed parameter space  
with a high confidence level. Even small NP contributions like 
$|C_1^{\prime}|$ and/or $|C_2^{\prime}|$ of order $10^{-1}$ can be differentiated from the SM 
at more than 5$\sigma$ confidence level. The point $C'_1=-1,\, C'_2=0$ denotes completely destructive 
interference with the SM and no signal events, and this is obviously much away from the SM expectation. 
The separations are expectedly worse for ${\cal L}_{\rm int}=2$ ab$^{-1}$, as shown in the right 
panels of the Figs. \ref{btkstnu} and \ref{btxsnu}, but even then, there are regions in the parameter 
space that can show some interesting trend. As an example, we note that it is possible to 
separate out NP contributions like 
$|C_1^{\prime}|, |C_2^{\prime}|$ $\approx 1$ from the SM at 5$\sigma$ confidence level or more. 

With enough data, one may even measure the differential decay distribution $d\Gamma/dq^2$.
In Figs. \ref{drbtkstnu} and \ref{drbtxsnu}, we show the differences in $d\Gamma/dq^2$ profiles 
between the SM and the NP for a couple of benchmark points, shown as NP-1 and NP-2, 
for the decays $B\to K^{\ast}\nu\bar\nu$ and $B\to X_s\nu\bar\nu$. 
Note that both the benchmark points are allowed even by the new Belle data (Fig.\ \ref{fig:c1c2}). 
Integrated branching fractions of these modes in SM and the selected benchmark points are listed in eq.(\ref{predvec}).
\begin{align}\label{predvec}
 \nn {\rm Br}(B\to K^*\nu\bar\nu)_{{\rm SM}} &= (9.43 \pm 1.48) \times 10^{-6}\,,\ \\
 \nn {\rm Br}(B\to K^*\nu\bar\nu)_{{\rm NP - 1}} &= (17.77 \pm 2.86) \times 10^{-6}\,,\ \\
 \nn {\rm Br}(B\to K^*\nu\bar\nu)_{{\rm NP - 2}} &= (3.99 \pm 0.70) \times 10^{-6}\,,\ \\
 \nn {\rm Br}(B\to X_s\nu\bar\nu)_{{\rm SM}} &= (28.88 \pm 1.90) \times 10^{-6}\,,\ \\
 \nn {\rm Br}(B\to X_s\nu\bar\nu)_{{\rm NP - 1}} &= (49.40 \pm 3.24) \times 10^{-6}\,,\ \\
     {\rm Br}(B\to X_s\nu\bar\nu)_{{\rm NP - 2}} &= (8.61 \pm 0.56) \times 10^{-6}\,.\ 
\end{align}

The present data almost rules out $|C'_i|\gg 1$ --- the NP has to be either loop-mediated or the 
new particles have to be so massive as to lie outside the direct detection range of the LHC --- and so 
we concentrate on small-$C'_i$ points. 
The bars shown in the plots represent the combined errors due to the various theory inputs, mostly coming 
from the form factors.
In the case of NP, if we treat the $\delta C'_i$s coming from the OO analysis 
of the respective decay modes as a measure of future statistical uncertainties on the NP WC's, then both $d Br/d q^2$ and the total 
branching fraction will have additional errors coming from them. 
We note that the NP sensitivities on the $q^2$ distributions of the exclusive and 
inclusive decay modes are different. For example, the distribution for
$B\to K^{\ast}\nu\bar\nu$ is highly sensitive to NP in the region $10~{\rm GeV}^2 <q^2 < 15~{\rm GeV}^2$, 
while that for the decay $B\to X_s\nu\bar\nu$ is more sensitive to the low-$q^2$ region, 
$q^2 < 10$ GeV$^2$. Therefore, study of $d\Gamma/dq^2$ for 
these exclusive and inclusive channels may be quite useful to pin down the parameters of the NP. 
For most of the beyond-SM theories, there should be a corroborative signature from charged lepton final 
state channels, but, as we pointed out, this may not be true always. 

Other observables are expected to yield different confidence level contours. This is shown in 
Fig. \ref{fig:tranpolneu} for the modified transverse polarization fraction $F'_T$, both for low and high 
${\cal L}_{\rm int}$. Note that the NP sensitivities of this observable are similar to that for the decay 
$B\to K^{\ast}\nu\bar\nu$. Note that the separation for this observable may not go beyond $3\sigma$ 
confidence level for the low-${\cal L}_{\rm int}$ option.

\section{Results: Invisible light scalars}

  \begin{figure*}[htbp]
\centering
\subfloat[\label{longms05} $m_S = 0.5$ GeV]
 {\includegraphics[height=5.5cm]{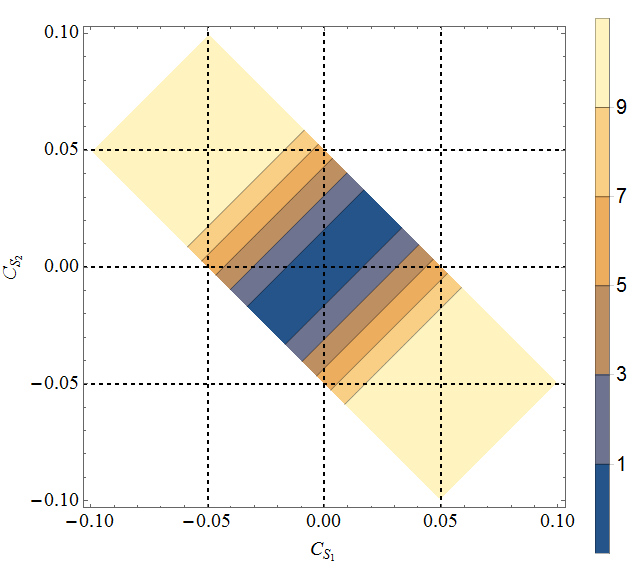}~
    \includegraphics[height=5.5cm]{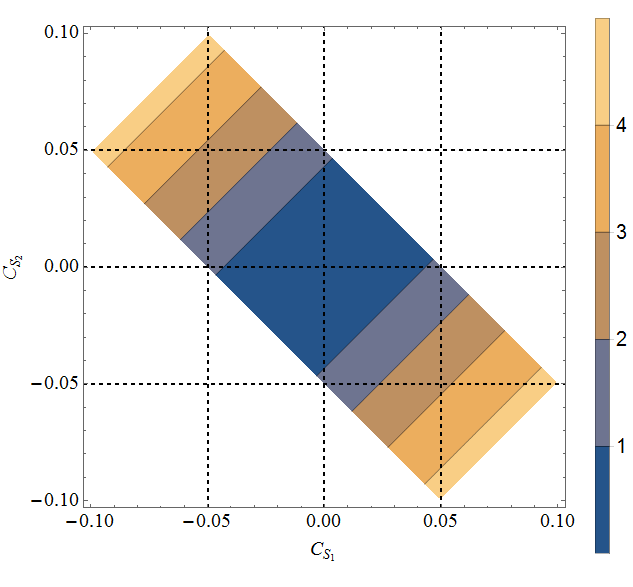}}\\
 \subfloat[\label{longms1.8} $m_S = 1.8$ GeV]  
{\includegraphics[height=5.5cm]{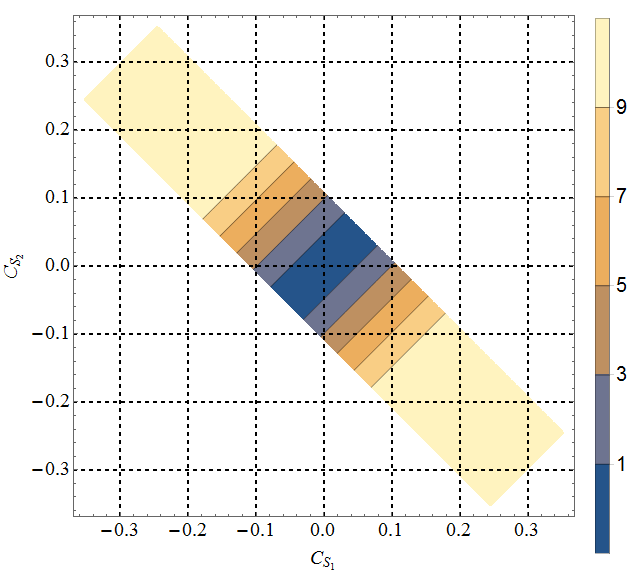}~
    \includegraphics[height=5.5cm]{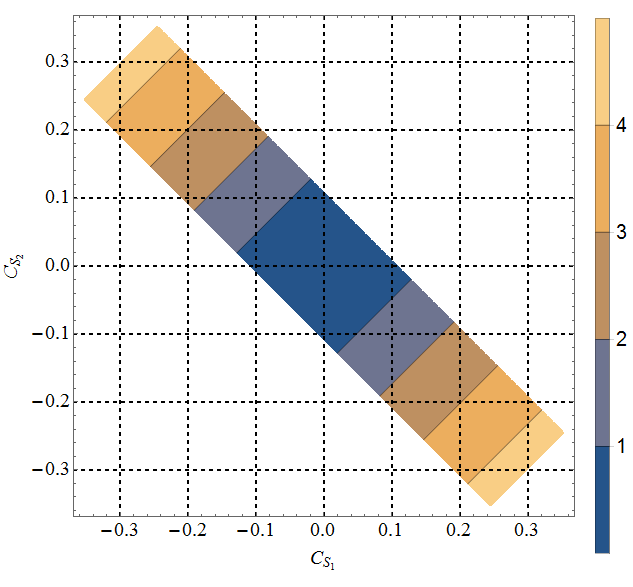}} 
 \caption{The contours from the measurement of $F'_L$ (Eq.\ (\ref{flmod})), with 
$m_S=0.5$ GeV and $1.8$ GeV, ${\cal L}_{\rm int}=50$ ab$^{-1}$ (left panels), 2 ab$^{-1}$ (right panels). 
}
\label{fig:longpol}
\end{figure*}
 
With invisible light scalars escaping the detector, one gets an identical signal as $b\to s\nu\bar\nu$. 
We will assume two such identical light scalars produced in the decay, {\em i.e.}, $b\to sSS$. The 
differential decay distributions depend on the mass of $S$, which we take to be either 0.5 GeV 
(called the light scalar or LS option), or 1.8 GeV (called the heavy scalar or HS option). For both these 
options, we show our results taking ${\cal L}_{\rm int} = 50$ ab$^{-1}$ and 2 ab$^{-1}$, just as before. 
Obviously, separation from the SM will be better for lighter scalars,
as for heavier scalars, the low-$q^2$ region will be covered only by the SM and hence those bins 
will be irrelevant for the analysis.  

We show the confidence levels in Figs.\ \ref{bkss}, \ref{bkstss} and \ref{bxsss} for the decays 
$B\to KSS$, $B\to K^{\ast}SS$ and $B\to X_s SS$ respectively for the LS option. 
The shape of the contours are intuitively obvious from the expressions of $d\Gamma/dq^2$. 
For example, the mode $B\to KSS$ is not of much use if $C_{S_1} \approx - C_{S_2}$. 
A complementary set of information can be obtained from 
the $B\to K^*SS$ mode. As expected, only regions close to the SM point $C_{S_1}=C_{S_2}=0$ may not be
differentiable from the SM itself.  Roughly speaking, one can have a $5\sigma$ separation from the SM 
in at least one channel with ${\cal L}_{\rm int}=50$ ab$^{-1}$ if
$|C_{S_1}|$ and/or $|C_{S_2}|$ be as small as 0.03. 
The inclusive channel $B\to X_s SS$ is even more powerful, as the branching fraction depends  
on the combination $|C_{S_1}|^2 + |C_{S_2}|^2$. This leads to circular contours around the origin. 
There is a subleading term proportional to the strange quark mass $m_s$ which breaks this symmetry, 
and so the contours appear to be slightly deformed. 
The low-luminosity option as displayed in the right panels show that the differentiation is harder 
for exclusive modes, while the inclusive mode fares better. 
Points like $|C_{S_1}|$ and/or $|C_{S_2}| \approx 0.05$ can be differentiated from the 
SM at more than 5$\sigma$ confidence level. Integrated branching fractions of these modes in SM and the selected 
benchmark point (fig. \ref{fig:drbsSS05}) are listed in eq.(\ref{predLS}).
\begin{align}\label{predLS}
 \nn {\rm Br}(B\to K + {\rm invis})_{{\rm SM}} &= (3.86 \pm 0.53) \times 10^{-6}\,,\ \\
 \nn {\rm Br}(B\to K + {\rm invis})_{{\rm NP - 1}} &= (8.38 \pm 0.75) \times 10^{-6}\,,\ \\
 \nn {\rm Br}(B\to K^* + {\rm invis})_{{\rm SM}} &= (9.43 \pm 1.48) \times 10^{-6}\,,\ \\
 \nn {\rm Br}(B\to K^* + {\rm invis})_{{\rm NP - 1}} &= (24.09 \pm 2.79) \times 10^{-6}\,,\ \\
 \nn {\rm Br}(B\to X_s + {\rm invis})_{{\rm SM}} &= (28.88 \pm 1.90) \times 10^{-6}\,,\ \\
     {\rm Br}(B\to X_s + {\rm invis})_{{\rm NP - 1}} &= (12.34 \pm 0.45) \times 10^{-5}\,.
\end{align}

Similar set of plots for the HS option are shown in Fig.\ 
\ref{fig:bsSS1.8}. The nature of the plots is identical to that of Fig.\ \ref{fig:bsSS05}. 
However, to reach the same sensitivity, one needs higher WCs than the LS option, as the NP effects are visible 
only in the low-$q^2$ bins. Integrated branching fractions of these modes for the selected 
benchmark point (fig. \ref{fig:drbsSS1.8}) are listed in eq.(\ref{predHS}).
\begin{align}\label{predHS}
 \nn {\rm Br}(B\to K + {\rm invis})_{{\rm NP - 1}} &= (5.69 \pm 0.52) \times 10^{-6}\,,\ \\
 \nn {\rm Br}(B\to K^* + {\rm invis})_{{\rm NP - 1}} &= (11.27 \pm 1.11) \times 10^{-6}\,,\ \\
     {\rm Br}(B\to X_s + {\rm invis})_{{\rm NP - 1}} &= (34.97 \pm 1.72) \times 10^{-6}\,.
\end{align}

The $q^2$ distributions for the decay rates of  $B\to KSS$, $B\to K^{\ast}SS$ and $B\to X_s SS$
are shown in Figs.\ \ref{drbkss}, \ref{drbkstss}, and \ref{drbxsss} respectively for ${\cal L}_{\rm int} = 50$ 
ab$^{-1}$ for the LS option. 
While for $B\to KSS$ and $B\to K^{\ast}SS$, the $q^2$ distributions are sensitive for NP over the 
entire $q^2$ region except for very high ($>15$ GeV$^2$) and very low ($\approx 0$) regions, 
for the semi-inclusive decay the NP sensitivity is more in the region $2~{\rm GeV}^2 < q^2 < 10~{\rm GeV}^2$.
Similar plots for $m_S=1.8$ GeV are shown in Fig. \ref{fig:drbsSS1.8}. We note that though 
$q^2_{\rm min}$ is much higher for this case, the nature of the $q^2$ distributions, and therefore the NP 
sensitivities, is similar to that obtained for the LS case. 

As is defined in Eq.\ \ref{flmod}, the decay $B\to K^*SS$ has another observable, namely, 
the longitudinal polarization $F'_L$. This is because the $K^*$ mesons appearing with scalars are 
completely longitudinally polarized.  The confidence level contours for $F'_L$ obtained from the OO 
analysis are shown in Figs.\ \ref{longms05} and \ref{longms1.8} for the LS and the HS options 
respectively. This observable has similar kind of NP sensitivity as that of $B\to K^{\ast} SS$.

\section{Summary}
We have analysed the NP sensitivities of the different observables in the decays $b\to s~+$~invisibles using 
the Optimal Observables technique. 
We consider two NP models: (1) only neutrinos as the carrier of missing energy but with a new 
operator involving right-handed quark current; and (2) apart from the SM neutrinos, light invisible scalars 
as the carrier of missing energy. The analysis takes into account all the new effective operators 
and their effects on several observables, namely, the total decay width for inclusive and exclusive modes, 
the differential decay distributions, and the modified transverse and longitudinal polarization fractions
as defined in the text. 

We show our results both for the high- and low-luminosity options of Belle-II, namely, 
${\cal L}_{\rm int} = 50$ ab$^{-1}$ and $2$ ab$^{-1}$ respectively. 
All the observables are sensitive to NP effects, and even
small NP effects might be detectable at future high-luminosity Belle-II. 
The differentiation of the NP from the SM is obviously not that trivial for the low-luminosity 
option, apart from the observables like inclusive branching fractions. 

The NP sensitivities of $d\Gamma/dq^2$ for exclusive and inclusive channels are different.
As the data on that will possibly come after the branching fraction data, they will serve as an additional 
check on the operator structure and parameter values of the NP. 
Note that the exclusive distributions are more or less similar for both the NP models, but the inclusive 
distributions are different, so that may serve as a good discriminator. 
Thus, we encourage our experimental colleagues to investigate both the $q^2$-integrated 
branching fractions as well as differential distributions. 


{\em Acknowledgements} -- Z.C.\ acknowledges the University of Calcutta for a research fellowship. 
A.K.\ acknowledges the Council for Scientific and Industrial Research, 
Government of India, for support through 
a research grant. He also acknowledges the hospitality of the Physics Department of IIT Guwahati, where 
some part of the work was completed.

\end{document}